\documentclass[useAMS,traditabstract]{aa}
\usepackage[dvips]{graphicx}
\usepackage{natbib}
\bibpunct{(}{)}{;}{a}{}{,} 
\usepackage{txfonts}
\usepackage{float}
\usepackage{rotating}

\newcommand{\mic}{\hbox{$\mu$m}}
\newcommand{\hii}{\mbox{H\,{\sc ii}}}

\newcommand{\tbgswarm}{\hbox{$T_\mathrm{W}^{\,\mathrm{BC}}$}}

\newcommand{\tbgscold}{\hbox{$T_\mathrm{C}^{\,\mathrm{ISM}}$}}
\newcommand{\fmu}{\hbox{$f_\mu$}}
\newcommand{\ldust}{\hbox{$L_{\mathrm{d}}^{\,\mathrm{tot}}$}}
\newcommand{\mdust}{\hbox{$M_{\mathrm{d}}$}}

\newcommand{\lsun}{\hbox{$\mathrm{L}_{\sun}$}}
\newcommand{\msun}{\hbox{$\mathrm{M}_{\sun}$}}

\newcommand{\xipahstot}{\hbox{$\xi_\mathrm{PAH}^\mathrm{\,tot}$}}

\newcommand{\ximirtot}{\hbox{$\xi_\mathrm{MIR}^\mathrm{\,tot}$}}

\newcommand{\xiwarmtot}{\hbox{$\xi_\mathrm{W}^\mathrm{\,tot}$}}

\newcommand{\xicoldtot}{\hbox{$\xi_\mathrm{C}^\mathrm{\,tot}$}}

\newcommand{\tauv}{\hbox{$\hat{\tau}_{V}$}}

\newcommand{\sfr}{\hbox{$\psi$}}
\newcommand{\ssfr}{\hbox{$\psi_{\mathrm S}$}}

\newcommand{\iras}{\hbox{\it IRAS}}
\newcommand{\galex}{\hbox{\it GALEX}}

\newcommand{\mstar}{\hbox{$M_{\ast}$}}

\begin{document}

\title{Exploring the physical properties of local star-forming ULIRGs from the ultraviolet to the infrared}

\author{E. da Cunha\inst{1,2}\thanks{e-mail: dacunha@physics.uoc.gr}
\and V. Charmandaris\inst{1,2,3}
\and T. D\'iaz-Santos\inst{1,2}
\and L. Armus\inst{4}
\and J. A. Marshall\inst{4,5}
\and D. Elbaz\inst{6}
}

\institute{Department of Physics, University of Crete, 71003 Heraklion, Greece
\and IESL/Foundation for Research and Technology-Hellas, 71110 Heraklion, Greece
\and Chercheur Associ\'e, Observatoire de Paris, 75014 Paris, France
\and Spitzer Science Center, California Institute of Technology, Pasadena, CA 91125, USA
\and The Jet Propulsion Laboratory, California Institute of Technology, Pasadena, CA 91125, USA
\and Laboratoire AIM, CEA/DSM-CNRS-Universi\'e Paris Diderot, IRFU/Service d'Astrophysique, B\^at.709, CEA-Saclay, 91191 Gif-sur-Yvette Cedex, France
}

\abstract{
We present an application of the da Cunha, Charlot \& Elbaz (2008) model of the spectral energy distribution (SEDs) of galaxies from the ultraviolet to far-infrared, to a small pilot sample of purely star-forming Ultra-Luminous Infrared Galaxies (ULIRGs). We interpret the observed SEDs of 16 ULIRGs using this physically-motivated model which accounts for the emission of stellar populations from the ultraviolet to the near-infrared and for the attenuation by dust in two components: an optically-thick starburst component and the diffuse ISM. The infrared emission is computed by assuming that all the energy absorbed by dust in these components is re-radiated at mid- and far-infrared wavelengths. This model allows us to derive statistically physical properties including star formation rates, stellar masses,  as well as temperatures and masses  of different dust components and plausible star formation histories.
We find that, although the ultraviolet to near-infrared emission represents only a small fraction of the total power radiated by ULIRGs, observations in this wavelength range are important to understand the properties of the stellar populations and dust attenuation in the diffuse ISM of these galaxies. Furthermore, our analysis indicates that the use of mid-infrared spectroscopy from the Infrared Spectrograph on the {\it Spitzer Space Telescope} is crucial to obtain realistic estimates of the extinction to the central energy source, mainly via the depth of the 9.7-$\mu$m silicate feature, and thus accurately  constrain the total energy balance.
Our findings are consistent with the notion that, in the local Universe, the physical properties of ULIRGs are fundamentally different from those of galaxies with lower infrared luminosities and that local ULIRGs are the result of merger-induced starbursts.
While these are well-established ideas, we demonstrate the usefulness of our SED modelling in deriving relevant physical parameters which provide clues to the star formation mode of galaxies.}


\titlerunning{The physical properties of star-forming ULIRGs}
\authorrunning{E. da Cunha et al.}

\maketitle

\section{Introduction} \label{intro}

Ultra-Luminous Infrared Galaxies (ULIRGs) are galaxies with infrared luminosities
higher than $10^{12}~L_{\sun}$. They emit most of their energy (over 90~per cent) in the infrared, which implies that they are heavily dust-obscured. To power their huge
infrared luminosities, these galaxies must be undergoing intense star formation,
forming new stars at rates of the order of $100~M_{\sun}$~yr$^{-1}$. In the local
Universe, such large starbursts are thought to result from major interactions or mergers of gas-rich galaxies (e.g.~\citealt{Murphy2001,Sanders1996} and references therein). This is supported by the fact that
local ULIRGs have typically disturbed morphologies (e.g.~\citealt{Clements1996,Duc1997,Rigopoulou1999}).
Another possible source of energy in ULIRGs is the accretion of large quantities of gas onto a central supermassive black hole in an active galactic nucleus (AGN; e.g.~\citealt{Sanders1988}). In such cases, dust in an optically-thick torus surrounding the AGN absorbs large amounts of energy and re-radiates it at longer wavelengths, contributing to the large infrared luminosities of ULIRGs. In most cases, there is likely to be a combination of both star formation and AGN contributing to the emission from ULIRGs, and the relative contribution by the two processes has been extensively investigated (e.g.~\citealt{Lutz1998,Genzel1998,Laurent2000,Armus2007,Spoon2007,Veilleux2009}). 

Understanding the formation and evolution of ULIRGs is key to understanding the cosmic evolution of galaxies.
ULIRGs play a significant role in galaxy evolution at high redshifts and they are major contributors to the infrared luminosity and star formation density at redshifts $z \sim1 - 2$ (e.g.~\citealt{Chary2001,LeFloch2005}). Local ULIRGs may also provide analogs to dusty galaxy populations at high redshifts (e.g.~\citealt{Blain2002,Daddi2007,Dey2008}).
To understand the evolution of ULIRGs in a cosmological context, the study of the physical properties of these system both at low and high redshifts is required. 

With the advent of a number of sensitive space observatories and new instrumentation on ground-based telescopes, a wealth of new data on local samples of ULIRGs have been compiled over recent years. The analysis of such data has focused mainly on the infrared range, where ULIRGs emit most of their energy. Several models have been proposed to interpret the infrared spectral energy distributions of ULIRGs (e.g.~\citealt{Klaas2001,Farrah2003,Vega2008,Marshall2007}). However, such models are not easily applicable in the statistical study of large samples of galaxies and they often neglect the optical emission of ULIRGs, which may carry important information about their intermediate-age and old stellar populations and hence their star formation histories.

Recently, \cite{daCunha2008} have presented a simple, physically motivated model that allows the interpretation of the integrated ultraviolet, optical and infrared spectral energy distributions (SEDs) of galaxies in terms of fundamental physical parameters. This model was used in \cite{daCunha2008} to derive median-likelihood estimates of parameters such as the star formation rate, stellar mass, dust attenuation and dust mass from the observed SEDs of 66 galaxies in the {\it Spitzer} Infrared Nearby Galaxy Survey (SINGS, \citealt{Kennicutt2003}). In another study, \cite{daCunha2009} have used this model to investigate the relation between star formation and dust content of a sample of 3258 low-redshift \iras\ galaxies detected by Sloan Digital Sky Survey (SDSS) with complementary observations by \galex\ (ultraviolet) and 2MASS (near-infrared). These studies have provided valuable insight into the physical properties of local galaxies with moderate total infrared luminosities ($L_\mathrm{IR} \lesssim 10^{12}~L_{\odot}$). 

The method of \cite{daCunha2008} was designed to interpret SEDs with sufficiently wide wavelength coverage (from the UV to the far-IR) to allow for accurate extinction corrections and energy balance. The main assumptions are that the energy input originates only from stars and all energy absorbed by dust in the ultraviolet and optical is re-radiated in the infrared. In this paper, we investigate how the model of \cite{daCunha2008} can be calibrated to interpret multi-wavelength observations of star-forming ULIRGs, which have much higher infrared luminosities and for which only a small fraction of the energy is detected in the ultraviolet and optical. This is the first step of a larger study in which we aim at expanding our methodology to interpret the physical properties of samples of galaxies covering a wide range of infrared luminosities, star formation and AGN activities, such as the Great Observatories All-Sky LIRG Survey (GOALS; \citealt{Armus2009}) and the local ULIRG sample of \cite{Desai2007}.

In the present study, we start by analysing the SEDs of 16 purely star-forming ULIRGs from the primary sample of \cite{Desai2007}. In a forthcoming paper, we will extend our analysis to galaxies with AGN contribution.

This paper is organized as follows. In Section~\ref{sample}, we describe our sample of local ULIRGs and the multi-wavelength data used in our study. We detail our method to derive statistical constraints on the physical parameters of our galaxies from fits to their observed spectral energy distributions in Section~\ref{method}. In Section~\ref{results}, we provide the fits to the SEDs of the galaxies and the estimates of their physical parameters. In Section~\ref{discussion} we show that our results are consistent with previous findings regarding the formation and evolution of ULIRGs, and compare the physical properties of local galaxies in two bins of total infrared luminosity: $10 \le \log(L_\mathrm{IR}/\lsun) < 12$ and $12 \le \log(L_\mathrm{IR}/\lsun) < 12.5$. Our summary and conclusions are presented in Section~\ref{conclusion}. When necessary, we adopt the following cosmology: $H_0=70$~km~s$^{-1}$~Mpc$^{-1}$, $\Omega_\mathrm{M}=0.30$ and $\Omega_\Lambda=0.70$.

\section{The sample} \label{sample}

We select our galaxies from a large sample of 107 local ULIRGs with $0.02 \lesssim z \lesssim 0.93$ observed in the $5-38~\mic$ region with the {\it Spitzer}/IRS \citep{Armus2007,Desai2007}. The sources were primarily selected from the \iras\ 1 Jy survey \citep{Kim1998}, the \iras\ 2 Jy survey \citep{Strauss1992} and the FIRST/\iras\ radio-far-infrared sample of \cite{Stanford2000}.
We note that preliminary theoretical interpretation of infrared observations of this sample was preformed by \cite{Armus2007} using the spectral decomposition method of \cite{Marshall2007}. 

\subsection{Selection of our star-forming ULIRGs}

\begin{figure}
\begin{center}
\includegraphics[width=0.4\textwidth]{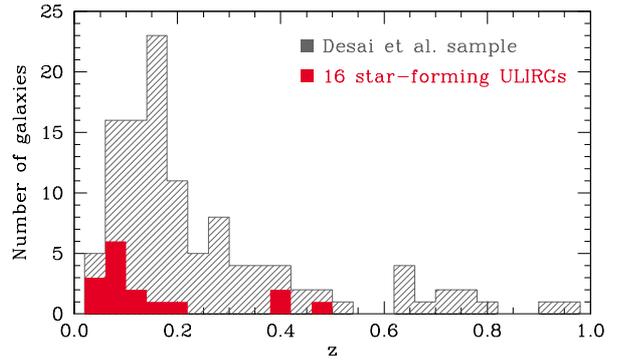}
\caption{Redshift distribution of the 16 star-forming ULIRGs used in this study (red histogram). For comparison, the grey histogram shows the redshift distribution of the primary sample of 107 local ULIRGs of \cite{Desai2007}.}
\label{fig1}
\end{center}
\end{figure}

Given the theoretical tools we have in hand, in the present study, our goal is to explore how  the physical properties of purely star-forming ULIRGs can be characterized using the model of \cite{daCunha2008}. Modelling the SEDs of ULIRGs presents a challenge since theses systems have infrared luminosities a factor of 10 greater than what had been studied so far \citep{daCunha2008,daCunha2009} and very high infrared-to-optical ratios. Typically more than 90\% of the energy output of ULIRGs is emitted in the infrared. This energy originates not only from young massive, heavily obscured starbursts, but in some cases from an AGN. Identifying the presence of an AGN in the dust-enshrouded nucleus of a ULIRG is often challenging, in particular in the optical, where the extinction effects are more prevalent. The best diagnostic is the detection of hard X-ray emission, or high ionization lines which cannot be excited by stellar photospheric emission. The task is somewhat easier in the infrared, and various methods have been developed to reveal an AGN and quantify its contribution to the bolometric emission of galaxy
(see e.g.~\citealt{Genzel1998, Laurent2000, Imanishi2006, Armus2007, Charmandaris2008,Veilleux2009}). In the present study,
we focus on the modelling of a sample of ULIRGs with no indication of significant AGN contribution. The inclusion of galaxies with AGN activity will be the subject of a forthcoming study.

We select the star-forming ULIRGs to be studied here from the primary sample of \cite{Desai2007} . Our selection process of the 16 systems is not designed to be complete. Following the motivations described above we have identified galaxies for which none of the well-known optical, near-infrared and mid-infrared spectroscopic diagnostics show indication of an AGN dominating in the optical or mid-infrared. Our 16 galaxies have strong  polycyclic aromatic hydrocarbon (PAH) emission, in particular at 6.2~\mic, and no [NeV]$\lambda 14.33\mic$ emission, or a mid-IR continuum indicative of an AGN. Additionally, all the ULIRGs in our sample can be classified as `cool' according to their \iras\ 25- to 60-\mic\ flux ratios (i.e. $F_{25}/F_{60} < 0.2$, see Table~\ref{tab:basic}), which has been taken as an indication of no significant AGN contribution to the mid-IR emission (e.g.~\citealt{Sanders1988,Farrah2007}). We note that even though two systems (IRAS12112+0305 and IRAS16334+4630) are classified as LINERs from optical spectroscopy in the literature \citep{Kimetal1998,Veilleux2002}, this does not necessarily imply the presence of an AGN but rather evidence of starburst-driven shocks \citep[see][]{Kimetal1998,Lutz1999}. Furthermore, detailed analysis of the infrared emission of these galaxies using the method of \cite{Marshall2007} shows that the contribution of an AGN to their infrared luminosity would be negligible (in the case of IRAS12112+0305, this is further supported by the SED analysis of \citealt{Farrah2003}, who estimate that an AGN would contribute by at most 0.1\% to the total infrared luminosity of this galaxy).

In Fig.~\ref{fig1}, we compare the redshift distribution of the primary sample of \cite{Desai2007} with that of our sub-sample of 16 star-forming ULIRGs. We note that our galaxies are at typically lower redshift than the bulk of the galaxies from the primary sample due to the availability of higher signal-to-noise data for nearby systems which allowed us to select them as purely star-forming. The basic properties
of our ULIRGs are listed in Table~\ref{tab:basic}.

\begin{table*}
\caption{Basic properties of the 16 ULIRGs analyzed in this paper. (a) \iras\ name; (b) right ascension (from NED); (c) declination (from NED); (d) redshift; (e) total infrared luminosity (8 -- 1000~\mic), computed using the \iras\ fluxes at 12, 25, 60 and 100~\mic\ (Table \ref{tab:data_fir}), following the prescription in \cite{Sanders1996}; (f) ratio of the \iras\ fluxes at 25 and 60~\mic.} \label{tab:basic}
\begin{center}
\begin{tabular}{lcccccc} \hline \hline
\\
Galaxy & R.A. (J2000) & Decl. (J2000) & $z$ & log($L_\mathrm{IR}/L_{\odot}$) & $F_{25}/F_{60}$ \\
(a) & (b) & (c) & (d) & (e)  & (f) \\
 \hline
IRAS00199-7426 & $00^\mathrm{h}22^\mathrm{m}07^\mathrm{s}.0$ & $-74^{\circ}09^{\prime}42^{\prime\prime}$ & $0.0963$ & $12.29$ & $0.078$ \\
IRAS01494-1845 & $01^\mathrm{h}51^\mathrm{m}51^\mathrm{s}.4$  & $-18^{\circ}30^{\prime}46^{\prime\prime}$ &$0.1579$ & $12.20$ & $0.059$ \\
IRASZ02376-0054 & $02^\mathrm{h}40^\mathrm{m}08^\mathrm{s}.6$ & $-00^{\circ}42^{\prime}04^{\prime\prime}$ & $0.4097$ & $12.39$ & $0.024$\\
IRAS04114-5117 & $04^\mathrm{h}12^\mathrm{m}44^\mathrm{s}.2$ & $-51^{\circ}09^{\prime}41^{\prime\prime}$ & $0.1245$ & $12.18$ & $0.035$\\
IRAS06009-7716 & $05^\mathrm{h}58^\mathrm{m}37^\mathrm{s}.0$ & $-77^{\circ}16^{\prime}39^{\prime\prime}$ & $0.1169$ & $12.00$ & $0.083$\\
IRASF08208+3211 & $08^\mathrm{h}23^\mathrm{m}54^\mathrm{s}.6$ & $+32^{\circ}02^{\prime}12^{\prime\prime}$ & $0.3957$ & $12.32$ & $0.024$\\
IRASF10156+3705 & $10^\mathrm{h}18^\mathrm{m}34^\mathrm{s}.5$ & $+36^{\circ}49^{\prime}52^{\prime\prime}$ & $0.4897$ & $12.46$ & $0.016$\\
IRAS10565+2448 & $10^\mathrm{h}59^\mathrm{m}18^\mathrm{s}.1$ & $+24^{\circ}32^{\prime}35^{\prime\prime}$ & $0.0431$ & $12.03$ & $0.105$\\
IRAS12112+0305 & $12^\mathrm{h}13^\mathrm{m}46^\mathrm{s}.0$ & $+02^{\circ}48^{\prime}38^{\prime\prime}$ & $0.0726$ & $12.27$ & $0.057$\\
IRAS13120-5453 & $13^\mathrm{h}15^\mathrm{m}06^\mathrm{s}.3$  & $-55^{\circ}09^{\prime}23^{\prime\prime}$  & $0.0307$ & $12.24$ & $0.073$\\
IRAS16334+4630 & $16^\mathrm{h}34^\mathrm{m}52^\mathrm{s}.6$  & $+46^{\circ}24^{\prime}53^{\prime\prime}$ & $0.1909$ & $12.37$ & $0.067$\\
IRAS17208-0014 & $17^\mathrm{h}23^\mathrm{m}21^\mathrm{s}.9$  & $-00^{\circ}17^{\prime}01^{\prime\prime}$ & $0.0430$ & $12.39$ & $0.050$\\
IRAS19297-0406 & $19^\mathrm{h}32^\mathrm{m}21^\mathrm{s}.2$  & $-03^{\circ}59^{\prime}56^{\prime\prime}$ & $0.0857$ & $12.38$ & $0.071$\\
IRAS19458+0944 & $19^\mathrm{h}48^\mathrm{m}15^\mathrm{s}.7$ & $+09^{\circ}52^{\prime}05^{\prime\prime}$ & $0.0999$ & $12.31$ & $0.064$\\
IRAS20414-1651 & $20^\mathrm{h}44^\mathrm{m}18^\mathrm{s}.2$ & $-16^{\circ}40^{\prime}16^{\prime\prime}$ & $0.0870$ & $12.16$ & $0.061$\\
IRAS22491-1808 & $22^\mathrm{h}51^\mathrm{m}49^\mathrm{s}.2$ & $-17^{\circ}52^{\prime}23^{\prime\prime}$ & $0.0772$ & $12.15$ & $0.097$\\
\hline
\end{tabular}
\end{center}
\end{table*}

\subsection{Mid-infrared Spitzer/IRS spectroscopy}

All ULIRGs in the sample of \cite{Desai2007} were observed in staring mode with both subslits of the Short-Low (SL) and Long-Low (LL) modules of the IRS. The final spectra have a spectral resolution typically between 60 and 120 over the wavelength range from 5 to 38~\mic. More detail on the IRS observations and data reduction can be found in \cite{Desai2007}.

\begin{figure}
\begin{center}
\includegraphics[width=0.5\textwidth]{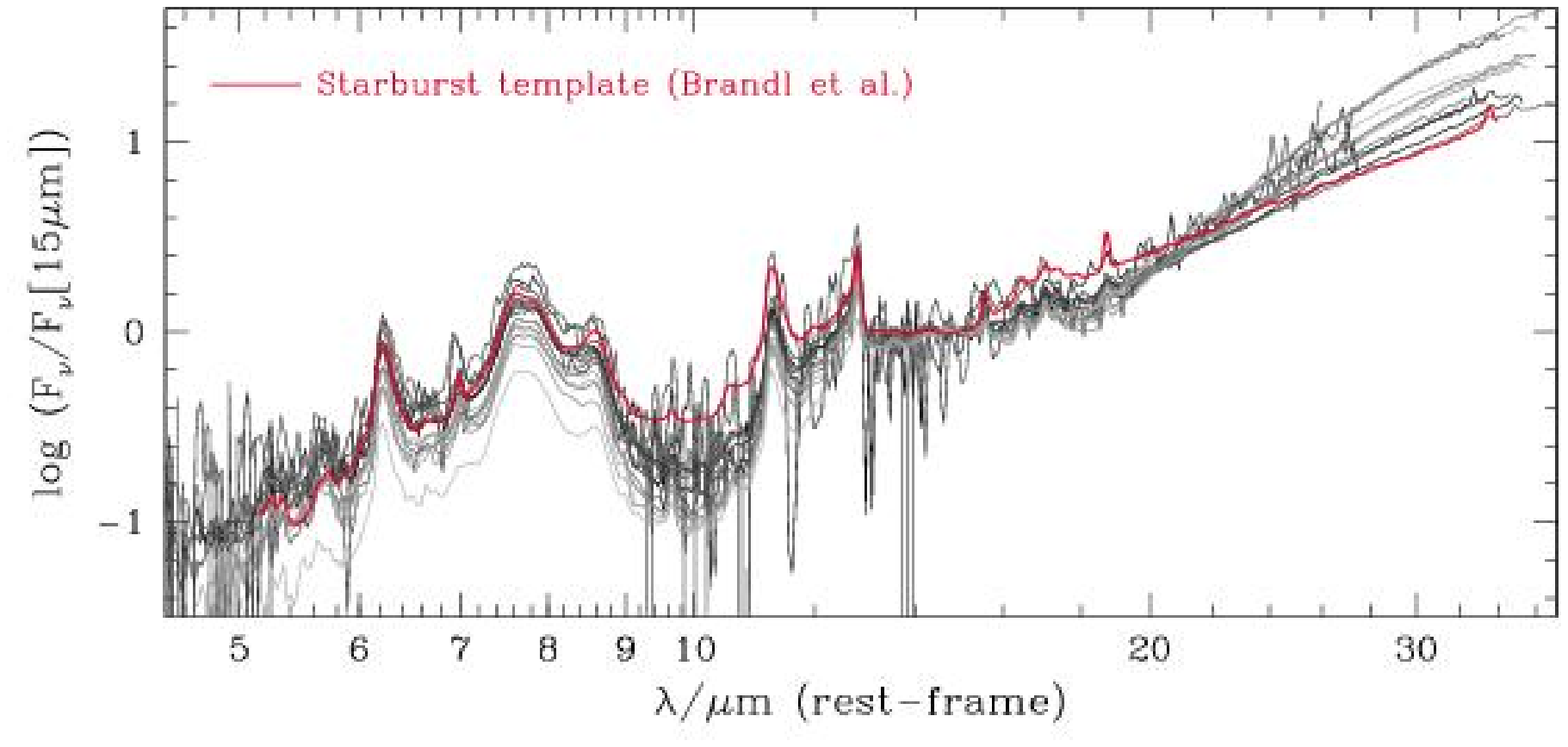}
\caption{IRS spectra of the 16 ULIRGs analyzed in this work (grey lines; \citealt{Desai2007,Armus2007}). These mid-infrared spectra show strong PAH features in the range 6 to 17~\mic, as well as a rising continuum from hot dust out to 35~\mic, indicating strong star formation in these systems. The strong absorption feature at $\lambda \simeq 10~\mic$ is due to silicates and indicates high dust optical depths in these galaxies. For comparison, we plot the starburst template of \cite{Brandl2006} in red. All the spectra are normalized to unity at 15~\mic. The strengths of the silicate absorption feature and the 6.2-\mic\ PAH equivalent widths of all spectra are presented in Table~\ref{tab:sil}.}
\label{fig2}
\end{center}
\end{figure}

In Fig.~\ref{fig2}, we show the mid-infrared IRS spectra of the 16 ULIRGs studied in this paper. The well-known PAH emission features, as well as silicate absorption, molecular hydrogen lines and neon fine structure lines are clearly visible.
It is interesting to compare the mid-infrared spectra of our ULIRGs with those of typical local starbursts of lower infrared luminosity. Therefore, we also plot the average starburst IRS spectrum of \cite{Brandl2006} in Fig.~\ref{fig2}. In general, the shapes of the mid-IR emission from our star-forming ULIRGs and the starburst template are remarkably similar, except for two striking differences: (i) most of our ULIRGs present much higher silicate absorption, meaning higher dust optical depths; (ii) the rising continuum from hot dust at
$\lambda \ga 20$~\mic\ is steeper for the ULIRGs, indicating that dust is heated to higher temperatures (see \citealt{Desai2007} for a more detailed analysis of the mid-infrared spectra of ULIRGs).

Given the small spatial scales in which star formation takes place, our ULIRGs are essentially unresolved (e.g.~\citealt{Soifer2001}, D\'iaz-Santos et al. 2010), and therefore our IRS spectra are representative of the total emission of the galaxies, and ca be used to characterize the global physical properties of our galaxies. We note that we do not aim at fitting the detailed shape of the mid-infrared spectra with the model used in this paper, but rather the global SEDs from the ultraviolet to the far-infrared. We use the IRS spectra to help select star-forming galaxies using mid-IR AGN diagnostics, and also to extract essential information such as the depth of the silicate feature at 9.7~\mic\ (Section~\ref{attenuation}) and mid-infrared broad-band colours (Section~\ref{photometry}) to provide general information about the shape of the mid-infrared continuum of the galaxies.

\subsection{Ancillary multi-wavelength photometry}\label{photometry}

We compile available photometric data in the ultraviolet, optical and infrared in order to obtain a sampling of the observed spectral energy distribution of each galaxy as complete as possible, which is crucial to fully exploit the potential of the \cite{daCunha2008} model. We include observations obtained with {\it GALEX}, SDSS, 2MASS, HST, {\it ISO}, \iras\ and {\it Spitzer} found in the NASA/IPAC Extragalactic Database (NED). Optical fluxes for IRAS17208-0014, IRAS19297-0406 and IRAS22491-1808 were kindly provided ahead of publication by D.~Sanders and V.~U. For consistency, all the selected fluxes correspond to the {\it total} flux from the galaxies in each band.
Additionally, we compute mid-infrared fluxes in the \iras\ 12-\mic\ and 25-\mic, {\it ISO} 15-\mic, {\it Spitzer}/IRAC 5.8-\mic\ and 8.0-\mic, and MIPS fluxes 24-\mic\ bands by convolving the observed IRS spectra with the respective filter response functions.

Several of our ULIRGs appear double in the optical and near-IR imaging (e.g.~\citealt{Veilleux2002}). However, in all cases we have {\it Spitzer} IRAC and/or IRS 16-\mic\ images which allow us to identify multiple mid-infrared emission components more clearly. These images reveal that, in the case of double systems, either the two components are too close and both are included within the 3.6~arcsec width of the IRS slits, or one of the two clearly dominates, contributing with over 90\% of the mid-infrared flux. We have checked that in these cases the IRS spectra are taken on the mid-infrared dominant source, which continues to remain unresolved at longer infrared wavelengths. In addition, for consistency, we use ultraviolet to near-infrared fluxes extracted using large apertures which include the total flux from both sources in the case of double systems.

In Appendix~\ref{data}, we list all the fluxes used in this work from the ultraviolet to the far-infrared.

\section{Method}\label{method}

\subsection{Spectral energy distribution modelling}\label{sed_mod}

We use the simple, physically motivated model of \citet{daCunha2008}
to interpret the mid- and far-infrared spectral energy
distributions of galaxies consistently with the emission at
ultraviolet, optical and near-infrared wavelengths.
This model was previously calibrated using samples of star-forming galaxies with 
moderate star formation rates and infrared luminosities.
Here we investigate how well this prescription can account for
the SEDs of local star-forming ULIRGs, which have much higher
infrared-to-optical ratios.
In this section, we briefly
recall the main features of the model of \cite{daCunha2008},
and describe the
modifications we have performed to the original
model in order to make it more suitable for the
interpretation of the multi-wavelength emission by ULIRGs.

\subsubsection{Stellar emission}

We compute the emission by stars in galaxies using the latest
version of the \citet{Bruzual2003} population synthesis code
(Charlot \& Bruzual, in preparation). This code predicts the
spectral evolution of stellar populations in galaxies from
far-ultraviolet to far-infrared wavelengths and at ages between
$1\times10^5$ and $2\times10^{10}$~yr, for different
metallicities, initial mass functions (IMFs) and star formation
histories. We adopt the \citet{Chabrier2003}
Galactic-disc IMF.

\subsubsection{Dust attenuation}\label{attenuation}

The ultraviolet, optical and near-infrared emission from stars is attenuated by dust.
We account for this using the
two-component model of \citet{Charlot2000}. This model includes
the fact that stars are born in dense molecular clouds with
finite lifetimes $t_{BC}$; after that time, these birth clouds dissipate as a result of 
the pressure caused mainly by strong stellar winds and supernovae explosions,
and stars migrate to the ambient (diffuse) ISM. Thus, the light produced by stars
younger than $t_{BC}$ is attenuated by dust in the birth clouds
and in the ambient ISM, while the light produced by older stars
is attenuated only by dust in the ambient ISM.
\cite{Charlot2000} have calibrated this model using a sample of local, ultraviolet-selected starburst galaxies
of low infrared luminosities, and show that a birth cloud lifetime of $t_{BC}=10^7$~yr can
account for the observed line and continuum emission properties of normal star-forming
galaxies.

Observations of local ULIRGs show that these galaxies are forming stars at 
high rates in a very concentrated central region with typical scale of at most a few kpc (e.g.
CO interferometer observations by \citealt{Bryant1996}, \citealt{Downes1998}; also
observations of Arp220 by \citealt{Sakamoto2008}). In this scenario,
the central region of the ULIRG would be similar to a huge, optically-thick molecular cloud with
a lifetime typically larger than $10^7$~yrs. Further evidence by \cite{Lahuis2007} suggests
the presence of high abundances of
warm, dense gas, associated with deeply embedded star formation in ULIRGs. In this environment,
\hii\ regions are prevented from expanding by large pressure gradients of gravity, thus
increasing the lifetime of the star formation process.
Therefore, in the framework of the attenuation model of
\cite{Charlot2000}, we adopt $t_{BC}=10^8$~yr for ULIRGs.
The interstellar medium is
still described using two components. The first is the `birth cloud' component,
which accounts for the central, optically-thick starburst powering most of the
infrared emission from these galaxies. The second component, the `diffuse ISM',
has smaller effective optical depths and is heated by stars older than $t_{BC}$. This
component is necessary to account for the small but still present
ultraviolet and optical emission from ULIRGs,
which cannot arise from the central starburst due to the extremely high optical depths
(see also spectroscopic evidence supporting this scenario in \citealt{Soto2010}).

We use an `effective absorption' curve for each
component, $\hat \tau_\lambda \propto \lambda^{-n}$, with the slope $n$
reflecting both the optical properties and the spatial distribution of the dust.
Following \citet{Charlot2000}, we adopt for the diffuse ISM
\begin{equation}
\hat\tau_\lambda^\mathrm{\,ISM}=\mu \hat\tau_V (\lambda/5500~\mathrm{\AA})^{-0.7}\,,
\label{tau_ism}
\end{equation}
where $\hat \tau_V$ is the total effective $V$-band absorption
optical depth of the dust seen by young stars inside birth clouds,
and $\mu = \hat\tau_V^\mathrm{\,ISM}/(\hat\tau_V^\mathrm{\,BC} +
\hat\tau_V^\mathrm{\,ISM})$ is the fraction of this contributed by
dust in the surrounding diffuse ISM. For the stellar birth clouds, we adopt:
\begin{equation}
\hat\tau_\lambda^\mathrm{\,BC} = (1-\mu) \hat\tau_V (\lambda/5500~\mathrm{\AA})^{-1.3}\,.
\label{tau_bc}
\end{equation}

The slopes of the effective absorption curve in the birth clouds and in the diffuse ISM are different
in order to account for the different spatial distributions of dust.
For the birth cloud component, we adopt $n=1.3$, which corresponds to a foreground dust screen.
This should be adequate to characterize the attenuation of optically-thick dust in the concentrated 
dense central starburst region of ULIRGs. The diffuse ISM is better described by a random distribution
of discrete dust clouds \citep{Charlot2000}, which can be accounted for adopting $n=0.7$
(i.e. a `greyer' effective absorption curve).

We use the prescription described above to compute the total energy absorbed by
dust in the `birth cloud' component (i.e. the dense central starburst) and in the surrounding `ambient ISM';
this energy is then re-radiated by dust at infrared wavelengths (assuming conservation of energy).
The total luminosity emitted by dust in the galaxy is then
\begin{equation}
L_\mathrm{d}^\mathrm{\,tot} = L_\mathrm{d}^\mathrm{\,BC} + L_\mathrm{d}^\mathrm{\,ISM} \,,
\label{ldust}
\end{equation}
where $L_\mathrm{d}^\mathrm{\,BC}$ and $L_\mathrm{d}^\mathrm{\,ISM}$ are the total
luminosity re-radiated by dust in the birth cloud and in the ambient ISM components,
respectively.

This approach, developed in \cite{Charlot2000} and \cite{daCunha2008}, allows us to interpret
the infrared emission from galaxies consistently with the emission at shorter wavelengths using a simple
energy balance argument (given that the main energy source in galaxies are stars, i.e. there is no contribution by an AGN).
For normal star-forming galaxies with moderate amounts of dust (optically-thin case), the effective $V$-band optical depth
\tauv\ seen by stars younger than $t_{BC}$ inside birth cloud is constrained by the attenuation suffered by emission lines,
e.g.~$\mathrm{H}\alpha/\mathrm{H}\beta$, while the effective $V$-band optical depth seen by older stars in the
diffuse ISM $\mu\tauv$ is mainly constrained by the attenuated ultraviolet and optical continuum emission from the
galaxy.

In starburst-dominated ULIRGs, practically all the radiation from stars in the birth cloud component is absorbed by the optically-thick dust,
and therefore it is not possible to reliably constrain \tauv\ using observations in the ultraviolet or optical. In such cases, where the
optical depths are very high, dust will absorb effectively even mid-infrared radiation, in particular in the silicate
band located around 9.7~\mic, where the grain cross-section is maximum (and also to a lesser extent at 18~\mic).
The {\it Spitzer}/IRS mid-infrared spectra of our ULIRGs allow us to measure the strength of the 9.7~\mic\ silicate feature for each one
of them. This feature strength can be converted to an apparent dust optical depth $\tau_{9.7}$, which can be used to compute the $V$-band optical depth
by assuming dust optical properties and geometry.
With this in mind, for each galaxy, we compute the strength of the 9.7-\mic\ silicate absorption feature, $S_{\mathrm{Si\,9.7\,\mu m}}$, from the observed IRS spectra, as:
\begin{equation}
\label{ee:siabs}
S_\mathrm{Si\,9.7\,\mu m}=\ln\frac{F^\mathrm{\,obs}_{\lambda}}{F^\mathrm{\,cont}_{\lambda}}\,,
\end{equation}
where $F^\mathrm{\,obs}_{\lambda}$ is the observed flux density of the feature and $F^\mathrm{\,cont}_{\lambda}$ that of the continuum, both evaluated at the wavelength, $\lambda_\mathrm{Si}$, where the absorption presents its maximum (typically about 9.7~\mic). To compute the $F^\mathrm{\,cont}_{\lambda}$ we fit the IRS spectrum of each galaxy to a simple linear function with anchors at 5.5 and 13.2~\mic\ (rest-frame) and interpolate it through the absorption feature.
This method is similar to that used by \cite{Sirocky2008} for measuring the silicate strength in PAH-dominated IRS spectra, but here we use a long-wavelength anchor at 13.2~\mic\ instead of at 14.5~\mic\ and a straight line instead of a power-law to fit the spectra. We note that both methods yield similar results for $S_{\mathrm{Si\,9.7\,\mu m}}$.
We have checked that our values of the silicate absorption strength, shown in Table~\ref{tab:sil}, are consistent with those found by \cite{Hao2007} for a larger sample of local ULIRGs and are significantly larger than the silicate absorption strength of local starbursts of lower infrared luminosity of \cite{Brandl2006}, as expected from Fig.~\ref{fig2}.

We assume that the observed silicate feature of our ULIRGs originates in the birth cloud component (which can be approximated by an optically-thick
dust screen), and use the ULIRG silicate model of \cite{Sirocky2008} to estimate the $V$-band optical depth  $\tau_{V}^{\,\mathrm{Si}}$
in the birth clouds from the apparent 9.7-\mic\ optical depth inferred from the silicate absorption feature strength.
The adopted model assumes a \cite{Mathis1977} grain size distribution and a cold \cite{Ossenkopf1992} dust screen, which
imply $\tau_{V}^{\,\mathrm{Si}} = 23.6\,\tau_{9.7}$, where $\tau_{9.7} \simeq -S_{\mathrm{Si\,9.7\,\mu m}}$ (see more details in \citealt{Sirocky2008}).
We include a 30\% uncertainty associated with $\tau_{V}^{\,\mathrm{Si}}$ derived
in this way to account for uncertainties related to the measurement of $S_{\mathrm{Si\,9.7\,\mu m}}$ (about 10\%) and model assumptions.
Table~\ref{tab:sil} shows the silicate absorption
feature strengths of our ULIRGs, and the $V$-band optical
depth predicted using the model of \cite{Sirocky2008}.
We will use these values to help constrain our model \tauv, as described in Section~\ref{fits}.

\begin{table}
\caption{Properties derived from the IRS spectra of our ULIRGs. (a) strength of the silicate absorption feature; (b) $V$-band optical depth computed using the ULIRG silicate model of \cite{Sirocky2008} assuming a cold dust screen; (c) equivalent width of the 6.2-\mic\ PAH feature\protect\footnotemark[1]. } \label{tab:sil}
\begin{center}
\begin{tabular}{lcccc} \hline \hline
\\
Galaxy & $S_{\mathrm{Si\,9.7\,\mu m}}$ & $\tau_{V}^{\,\mathrm{Si}}$ & EW$_{6.2}$/\mic\ \\
 & (a) & (b) & (c)  \\
 \hline
IRAS00199-7426 &  $-1.149$ & $27.11$ & 0.377 \\
IRAS01494-1845  &  $-1.396$ & $32.95$ & 0.394\\
IRASZ02376-0054  &  $-0.697$ & $16.44$ & 0.332\\
IRAS04114-5117  &  $-1.500$ & $35.40$ & 0.345\\
IRAS06009-7716  &  $-1.262$ & $29.78$ & 0.429\\
IRASF08208+3211  &  $-1.424$ & $33.60$ & 0.346\\
IRASF10156+3705  &  $-1.072$ & $25.30$ & 0.659\\
IRAS10565+2448  &  $-1.139$ & $26.89$ & 0.429\\
IRAS12112+0305  &  $-1.629$ & $38.43$ & 0.289\\
IRAS13120-5453  &  $-1.475$ & $34.82$ & 0.373\\
IRAS16334+4630 &  $-1.367$ & $32.27$ & 0.349\\
IRAS17208-0014  &  $-1.856$ & $43.80$ & 0.329\\
IRAS19297-0406  &  $-1.600$ & $37.77$ & 0.317\\
IRAS19458+0944 &  $-1.810$ & $42.71$ & 0.414\\
IRAS20414-1651  &  $-1.705$ & $40.23$ & 0.423\\
IRAS22491-1808 &  $-1.624$ & $38.33$ & 0.382\\
\hline
\end{tabular}
\end{center}
\end{table}

\footnotetext[1]{We measure the 6.2-\mic\ PAH feature equivalent width by, firstly, fitting the continuum emission close to the feature with a linear function anchored at 5.8 and 6.6~\mic. Then, compute the feature emission by integrating the continuum-subtracted spectrum from 5.9 to 6.5~\mic, and finally we divide this by the continuum emission interpolated at 6.2~\mic.}

\subsubsection{Dust emission}

We distribute $L_\mathrm{d}^\mathrm{\,BC}$ and $L_\mathrm{d}^\mathrm{\,ISM}$
in wavelength between 3 and 1000~\mic\ using four main components
(see \citealt{daCunha2008} for details):
(i) emission from polycyclic aromatic hydrocarbons (PAHs;
i.e.~mid-infrared emission features),
(ii) mid-infrared
continuum emission from hot dust with temperatures in the range
130--250~K,
(iii) emission from warm dust in thermal
equilibrium with adjustable temperature, and
(iv) emission from cold dust in thermal equilibrium with
adjustable temperature.

In stellar birth clouds, the relative contributions to
$L_\mathrm{d}^\mathrm{\,BC}$ by PAHs, the hot mid-infrared 
continuum and warm dust are kept as adjustable parameters.
These clouds are assumed not to contain any cold dust. In the
ambient ISM, the contribution to $L_\mathrm{d}^\mathrm{\,ISM}$
by cold dust is kept as an adjustable parameter. The relative
proportions of the other three components are fixed to the 
values reproducing the mid-infrared cirrus emission of the Milky
Way (more detail in section~2.2 of \citealt{daCunha2008}).

\subsection{Median-likelihood estimates of physical parameters}

We use the model described in the previous section to interpret the observed spectral energy distributions
of our ULIRGs and derive median-likelihood estimates of their physical parameters.
To do so, we adopt a Bayesian
approach similar to that used by \citealt{daCunha2008}
(see also \citealt{Kauffmann2003b}, \citealt{Brinchmann2004}, \citealt{daCunha2009}).
This method relies on a large library of random models encompassing all plausible parameter
combinations (e.g. star formation histories,
metallicities, dust optical depths, dust masses, dust temperatures).
Such a library was already built in \cite{daCunha2008,daCunha2009} to derive statistical constraints on the
physical parameters of local star-forming galaxies of moderate infrared luminosities. For each galaxy,
we build the likelihood distribution of any physical parameter
by evaluating how well each model in the library accounts for the observed SED of the galaxy.
The underlying assumption of our method is that the model library represents the distribution from
which the data were randomly drawn, so it is important that the prior distributions of the parameters
sample well the observational 
space and do not give too much weight to a priori implausible corners of the parameter space.

To reproduce the observations of our ULIRGs, we find that some of the parameter priors have to be changed
with respect to the priors used to model galaxies with $L_\mathrm{IR} \lesssim 10^{12}$~\lsun\
in \cite{daCunha2008,daCunha2009}. This confirms the idea that ULIRGs are probing a very different
region of the parameter space of galaxies.
In particular, we include higher dust optical
depths and temperatures and increase the probability and strengths of bursts of star formation. We describe
the priors in detail in the next section.

\subsubsection{Model library}\label{library}

We first build a large library of 50 000 stellar population models with random star formation histories, metallicities and
dust contents. We distribute the galaxy age uniformly between 2 and 13.5~Gyr.
We parametrize the star formation history of each model in terms of two components: an exponentially declining
star formation rate of the form $\exp(-\gamma t)$, where we distribute the time-scale parameter $\gamma$ between 0 and 
1 in the same way as \cite{daCunha2008}, and superimpose random bursts. 
Since our ULIRGs are highly star-forming galaxies, likely to be experiencing star formation bursts, we assign a 75 per cent probability
of a burst occurring in the past 2~Gyr. The amplitude of each burst is defined as the ratio between the mass of stars formed during the burst and the total
mass of stars formed by the underlying continuous model, $A=M_\mathrm{burst}/M_\mathrm{cont}$.
We distribute $A$ logarithmically between 0.1 and 10. The duration of each burst, $t_\mathrm{burst}$, is distributed uniformly between
$3\times10^7$ and $3\times10^8$~yr.
The metallicity is uniformly distributed between 0.2 and 2 times solar.

To account for the high optical depths of ULIRGs, we have to change significantly the priors of the attenuation parameters
\tauv\ (effective $V$-band optical depth seen by stars younger than $t_{BC}$ in the birth cloud component) and $\mu$ (fraction
of \tauv\ contributed by the diffuse ISM component), compared to the previous study of \cite{daCunha2008}. This is motivated
by the very high infrared-to-optical ratios observed in the galaxies for which optical observations are available, and also by the high
$V$-band optical depths $\tau_V^\mathrm{Si}$ derived from the IRS spectra using the silicate feature (Table~\ref{tab:sil}). 
The strength of the silicate feature in our ULIRGs implies that the birth cloud component has to be optically-thick,
and the fact that ultraviolet and/or optical
emission is observed for several of our galaxies means that the diffuse ISM component should have much smaller optical depths.
Therefore, we distribute \tauv\ uniformly in log between 2 and 50. These high optical depths imply that the optical emission from ULIRGs
comes from a more diffuse, optically-thin component -- the diffuse ISM component. An optically-thin ISM component is the only way
the optical fluxes can be fitted consistently without the diffuse ISM contributing too much to the total infrared luminosity (i.e. too high $\fmu=L_\mathrm{d}^\mathrm{\,ISM}/\ldust$). This is the motivation for drawing the $V$-band optical depth in the diffuse ISM, $\mu\tauv$ from a Gaussian prior centered at 1 with 0.5 dispersion, between 0 and 2.
We note that this is in agreement with studies of the extinction in spatially-resolved ULIRGs \citep{Garcia2009}, where values of optical extinction as low as 0.2~mag are
found for the external regions (in typical scales of 1 -- 3 kpc).

We also generate a library of 50 000 infrared emission models in a similar way to \cite{daCunha2008}. We distribute the fraction of
total dust luminosity contributed by the diffuse ISM, $\fmu=L_\mathrm{d}^\mathrm{\,ISM}/\ldust$, uniformly between 0 and 1.
We randomly draw the fractional contributions by PAHs, hot mid-infrared continuum and warm dust in thermal
equilibrium to the total infrared luminosity of birth clouds, from the same prior distributions as \cite{daCunha2008}.
The equilibrium temperature of warm dust in birth clouds, \tbgswarm, is distributed uniformly between 30 and 60~K.
For the equilibrium temperature of cold dust in the diffuse ISM (\tbgscold), we extend the prior of \cite{daCunha2008} to include higher values in order to
account for the observed hotter far-infrared colours of our ULIRGs; \tbgscold\ is uniformly distributed between 15 and 30~K.

Finally, we combine the library of stellar population models and the library of dust emission models by joining models with similar
\fmu\ parameter and scaling to the total infrared luminosity \ldust\ as detailed in \cite{daCunha2008}. For each model spectrum in our library,
we compute the synthetic photometry in photometric bands from the ultraviolet to the far-infrared at the redshifts of our ULIRGs.

\subsubsection{Spectral fits}\label{fits}

We perform spectral fits by comparing the observed spectral energy distributions and also the $V$-band optical depths of the birth cloud component
derived from the silicate strengths measured using the {\it Spitzer}/IRS spectra in Section~\ref{attenuation} to every model in our stochastic library.
Specifically, for each ULIRG, we compute the $\chi^2$ goodness-of-fit of each model $j$ in the library:
\begin{equation}
\chi_{j}^{2}=\sum_{i}^{} \left(\frac{L_{\nu}^{i}- w_j 
\times L_{\nu,j}^{i}}{\sigma_{i}}\right)^{2} + \left(\frac{\tau_V^\mathrm{Si}- \tauv_{,j}}{\sigma_\mathrm{Si}}\right)^{2}\,,
\label{chi2}
\end{equation}
where $L_{\nu}^{i}$ and $L_{\nu,j}^{i}$ are the luminosities in the 
$i^{\rm th}$ band of the observed galaxy and the $j^{\rm th}$ model, 
respectively, $\sigma_{i}$ is the observational uncertainty in 
$L_{\nu}^{i}$, $w_j$ is the model scaling factor that minimises $\chi^2$ [see eq.~(33) of \cite{daCunha2008}],
$\tau_V^\mathrm{Si}$ is the $V$-band optical depth derived from the observed 9.7~\mic\ silicate feature and
$\sigma_\mathrm{Si}$ is the uncertainty associated with this measure. As discussed in Section~\ref{attenuation},
we explicitly fit $\tau_V^\mathrm{Si}$ to the effective $V$-band optical depth in the birth clouds because they are
so optically-thick this information is not accessible from ultraviolet and optical observations.

We fit all the available observed broad-band fluxes of our galaxies from the ultraviolet to the far-infrared, plus the $V$-band
optical depth of the birth clouds inferred from the mid-infrared silicate absorption strength (eq.~\ref{chi2}). We do not
attempt at reproducing the detailed shape of the observed IRS spectra of our galaxies because our SED model is
not designed to reproduce the detailed emission in this spectral range (so that the number of free
parameters is kept minimal). However, we aim at reproducing the rough shape of the mid-infrared SED of our galaxies, since this
reflects the emission by hot, stochastically-heated dust, and also by PAHs. To do so, we compute
broad-band fluxes in the mid-infrared from the observed IRS spectra and include them as observables in the fitting procedure.

We build the likelihood distribution of any given physical parameter for our observed galaxies by weighting the value of
that parameter by the probability $\exp(-\chi^2/2)$. Our best estimate of the parameter is the median of the resulting 
probability density function and our confidence interval the 16th--84th percentile range.
The results of our SED fitting for the sample of 16 star-forming ULIRGs using this method are presented in the next section.

\section{UV-to-IR SEDs and the physical parameters of star-forming ULIRGs}\label{results}

\begin{table*}
\caption{Median-likelihood estimates and confidence ranges (computed as the 16th -- 84th percentile range of the likelihood distribution) of some relevant parameters derived from the multi-wavelength fits to the observed SEDs of our ULIRGs: (a) fraction of total dust luminosity contributed by the diffuse ISM; (b) specific star formation rate; (c) stellar mass; (d) dust luminosity; (e) dust mass; (f) equilibrium temperature of warm dust in birth clouds; (g) contribution of warm dust in thermal equilibrium to the total dust luminosity; (h) equilibrium temperature of cold dust in the diffuse ISM; (i) contribution of cold dust in thermal equilibrium to the total dust luminosity.} \label{tab:results}
\begin{center}
\begin{tabular}{lccccccccc} \hline \hline
 & & & & & & & & & \\
Galaxy & \fmu\ & $\log(\ssfr/M_{\odot}$~yr$^{-1}$) & $\log(\mstar/M_{\odot})$ & $\log(\ldust/L_{\odot})$  & $\log(\mdust/M_{\odot})$ & \tbgswarm\ & \xiwarmtot\ & \tbgscold\ & \xicoldtot\ \\
 & (a) & (b) & (c) & (d) & (e) & (f) & (g) & (h) & (i) \\
 \hline
\\[-0.75ex]
IRAS00199-7426 & $0.116^{+0.066}_{-0.127}$ & $-8.57^{+0.30}_{-0.35}$ & $10.82^{+0.27}_{-0.19}$ & $12.29^{+0.04}_{-0.02}$ & $8.74^{+0.38}_{-0.19}$ & $41^{+10}_{-3}$ & $0.781^{+0.059}_{-0.064}$ & $23^{+4}_{-5}$ & $0.113^{+0.071}_{-0.072}$\\[0.75ex]
IRAS01494-1845 & $0.132^{+0.034}_{-0.092}$ & $-8.28^{+0.05}_{-0.15}$ & $10.54^{+0.15}_{-0.07}$ & $12.24^{+0.04}_{-0.02}$ & $8.56^{+0.69}_{-0.27}$ & $54^{+2}_{-1}$ & $0.775^{+0.062}_{-0.038}$ & $22^{+2}_{-5}$ & $0.148^{+0.037}_{-0.052}$\\[0.75ex]
IRASZ02376-0054 & $0.096^{+0.653}_{-0.107}$ & $-8.93^{+0.30}_{-0.95}$ & $10.98^{+0.31}_{-0.12}$ & $12.45^{+0.02}_{-0.02}$ & $8.62^{+0.03}_{-0.04}$ & $55^{+3}_{-11}$ & $0.831^{+0.102}_{-0.617}$ & $18^{+4}_{-3}$ & $0.121^{+0.346}_{-0.085}$\\[0.75ex]
IRAS04114-5117 & $0.047^{+0.071}_{-0.004}$ & $-8.02^{+0.10}_{-0.15}$ & $10.39^{+0.48}_{-0.33}$ & $12.22^{+0.01}_{-0.01}$ & $8.56^{+0.02}_{-0.30}$ & $50^{+2}_{-0}$ & $0.893^{+0.000}_{-0.035}$ & $21^{+5}_{-0}$ & $0.072^{+0.035}_{-0.000}$\\[0.75ex]
IRAS06009-7716 & $0.088^{+0.069}_{-0.082}$ & $-8.18^{+0.00}_{-0.40}$ & $10.18^{+0.10}_{-0.01}$ & $12.00^{+0.02}_{-0.01}$ & $8.05^{+0.25}_{-0.21}$ & $55^{+1}_{-3}$ & $0.784^{+0.056}_{-0.037}$ & $25^{+3}_{-5}$ & $0.089^{+0.041}_{+0.038}$\\[0.75ex]
IRASF08208+3211 & $0.112^{+0.091}_{-0.099}$ & $-8.53^{+0.20}_{-0.25}$ & $10.90^{+0.22}_{-0.16}$ & $12.44^{+0.04}_{-0.07}$ & $8.64^{+0.28}_{-0.20}$ & $58^{-1}_{-1}$ & $0.831^{+0.078}_{-0.065}$ & $21^{+5}_{-2}$ & $0.128^{+0.049}_{-0.071}$\\[0.75ex]
IRASF10156+3705 & $0.104^{+0.165}_{-0.133}$ & $-9.02^{+0.80}_{-0.60}$ & $11.49^{+0.59}_{-0.69}$ & $12.58^{+0.05}_{-0.08}$ & $8.43^{+0.20}_{-0.24}$ & $54^{+2}_{-3}$ & $0.861^{+0.088}_{-0.035}$ & $28^{+1}_{-7}$ & $0.078^{+0.051}_{-0.065}$\\[0.75ex]
IRAS10565+2448 & $0.154^{+0.075}_{-0.106}$  & $-8.83^{+0.40}_{-0.20}$ & $10.63^{+0.36}_{-0.26}$ & $11.97^{+0.01}_{-0.00}$ & $8.13^{+0.25}_{-0.04}$ & $55^{+1}_{-1}$ & $0.652^{+0.074}_{-0.010}$ & $26^{+0}_{-3}$ & $0.205^{+0.007}_{-0.048}$\\[0.75ex]
IRAS12112+0305 & $0.048^{+0.000}_{-0.017}$ & $-7.98^{+0.00}_{-0.00}$ & $10.34^{+0.00}_{-0.00}$ & $12.33^{+0.00}_{-0.00}$ & $8.28^{+0.00}_{-0.00}$ & $48^{+0}_{-0}$ & $0.868^{+0.000}_{-0.000}$ & $27^{+0}_{-0}$ & $0.078^{+0.000}_{-0.000}$\\[0.75ex]
IRAS13120-5453 & $0.206^{+0.000}_{-0.155}$ & $-8.33^{+0.00}_{-0.00}$ & $10.49^{+0.00}_{-0.01}$ & $12.21^{+0.00}_{-0.01}$ & $8.88^{+0.00}_{-0.00}$ & $52^{+1}_{-0}$ & $0.672^{+0.065}_{-0.000}$ & $22^{+0}_{-1}$ & $0.247^{+0.000}_{-0.062}$\\[0.75ex]
IRAS16334+4630 & $0.110^{+0.025}_{-0.073}$ & $-8.47^{+0.20}_{-0.35}$ & $10.92^{+0.23}_{-0.21}$ & $12.43^{+0.01}_{-0.02}$ & $8.41^{+0.28}_{-0.05}$ & $57^{+2}_{-1}$ & $0.804^{+0.033}_{-0.001}$ & $26^{+1}_{-4}$ & $0.123^{+0.016}_{-0.028}$\\[0.75ex]
IRAS17208-0014 & $0.099^{+0.006}_{-0.057}$ & $-9.26^{+0.50}_{-0.35}$ & $11.13^{+0.12}_{-0.14}$ & $12.25^{+0.03}_{-0.02}$ & $8.53^{+0.19}_{-0.11}$ & $52^{+2}_{-0}$ & $0.849^{+0.000}_{-0.027}$ & $23^{+2}_{-2}$ & $0.103^{+0.020}_{-0.001}$\\[0.75ex]
IRAS19297-0406 & $0.090^{+0.010}_{-0.044}$ & $-8.82^{+0.15}_{-0.10}$ & $11.19^{+0.03}_{-0.14}$ & $12.37^{+0.01}_{-0.02}$ & $8.44^{+0.20}_{-0.14}$ & $54^{+0}_{-2}$ & $0.786^{+0.042}_{-0.000}$ & $25^{+3}_{-3}$ & $0.138^{+0.009}_{-0.042}$\\[0.75ex]
IRAS19458+0944 & $0.082^{+0.099}_{-0.092}$ & $-8.43^{+0.20}_{-0.50}$ & $10.66^{+0.35}_{-0.21}$ & $12.19^{+0.03}_{-0.03}$ & $8.21^{+0.74}_{-0.23}$ & $53^{+1}_{-1}$ & $0.851^{+0.041}_{-0.081}$ & $23^{+4}_{-6}$ & $0.066^{+0.098}_{-0.038}$\\[0.75ex]
IRAS20414-1651 & $0.061^{+0.033}_{-0.031}$ & $-8.34^{+0.25}_{-0.35}$ & $10.42^{+0.35}_{-0.13}$ & $12.20^{+0.02}_{-0.02}$ & $8.31^{+0.72}_{-0.44}$ & $52^{+1}_{-1}$ & $0.886^{+0.004}_{-0.029}$ & $23^{+7}_{-6}$ & $0.075^{+0.021}_{-0.006}$\\[0.75ex]
IRAS22491-1808 & $0.059^{+0.053}_{-0.058}$ & $-8.40^{+0.45}_{-0.05}$ & $10.62^{+0.07}_{-0.38}$ & $12.12^{+0.04}_{-0.02}$ & $7.91^{+0.17}_{-0.10}$ & $58^{+1}_{-1}$ & $0.882^{+0.040}_{-0.049}$ & $27^{+3}_{-5}$ & $0.077^{+0.047}_{-0.047}$\\[0.75ex]
\hline
\end{tabular}
\end{center}
\end{table*}

\begin{figure*}
\begin{minipage}[t]{0.5\linewidth}
\centering
\includegraphics[width=\textwidth]{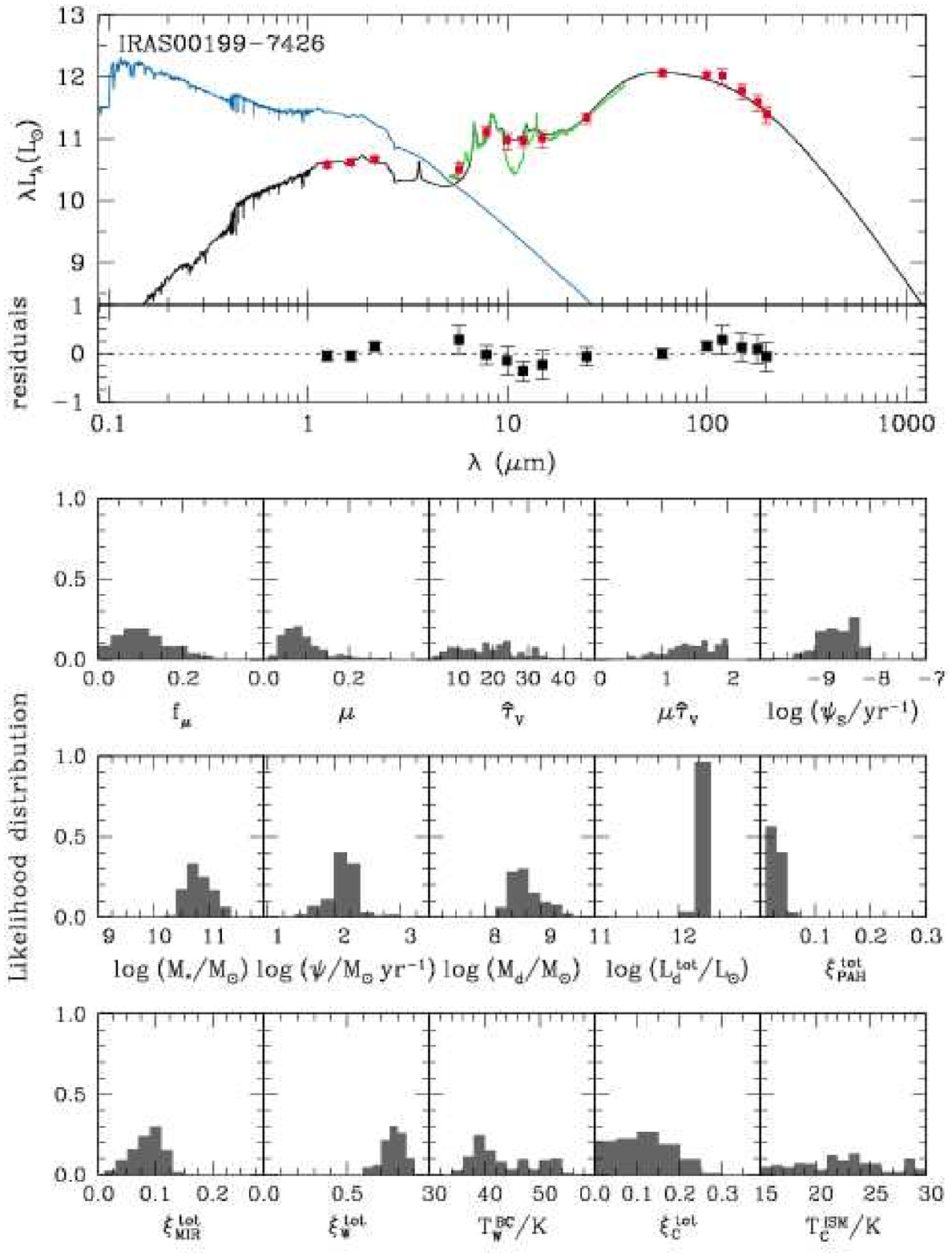}
\end{minipage}
\begin{minipage} [t] {0.5\linewidth}
\centering
\includegraphics[width=\textwidth]{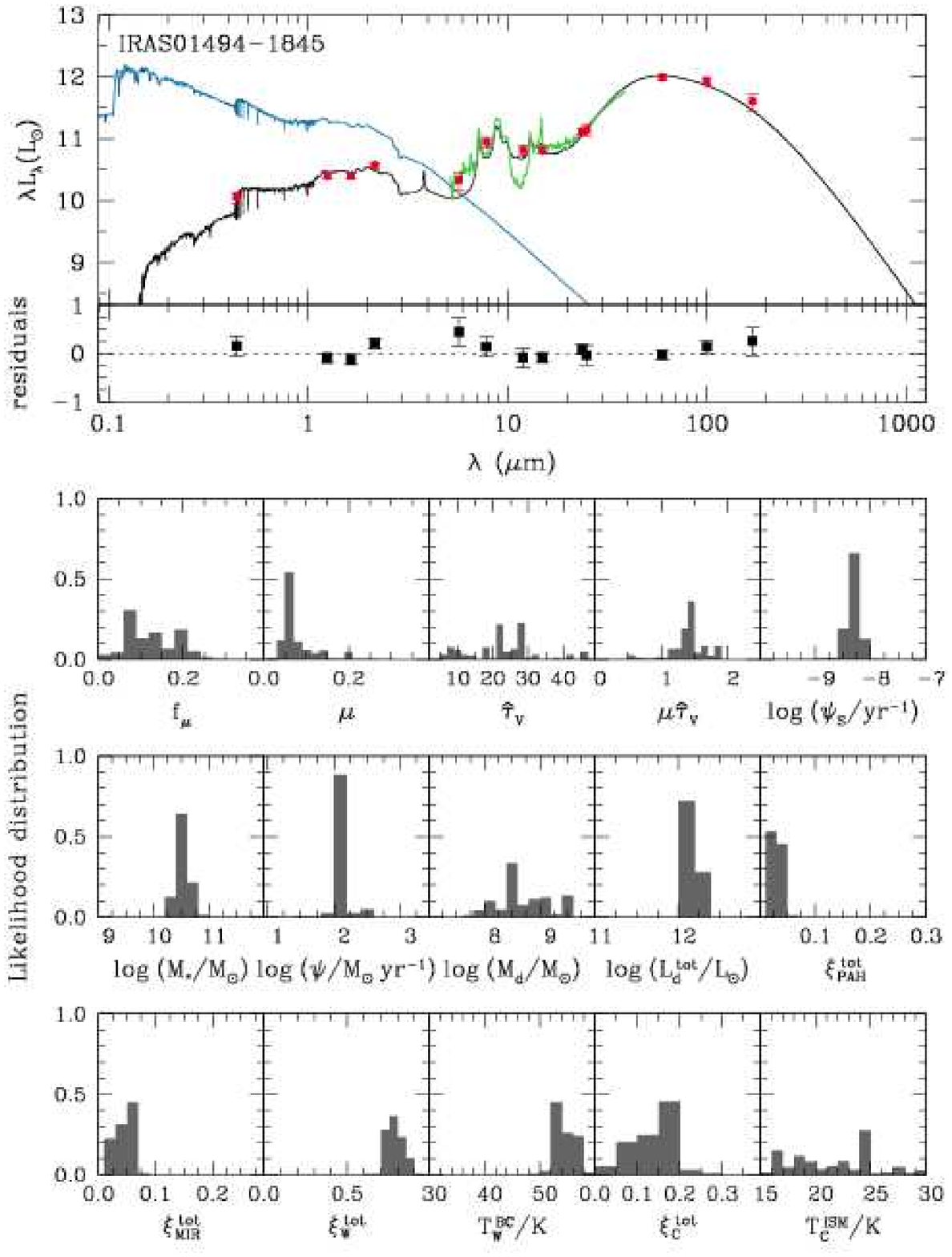}
\end{minipage}
\caption{{\it Top panels:} Best-fit models of the ULIRGs in our sample (observed-frame). The \iras\ name of the galaxy is indicated in the top left-hand corner. The blue line shows 
the unattenuated stellar spectrum, and the black line shows the total SED including the
attenuated stellar spectrum and the emission from dust in the infrared. The red squares are the
available observed broad-band luminosities, and the green line is the observed {\it Spitzer}/IRS spectrum of this galaxy. The residuals 
$(L_\lambda^\mathrm{obs}-L_\lambda^\mathrm{mod})/
L_\lambda^\mathrm{obs}$ are shown at the 
bottom of the panel.
{\it Bottom panels:} Likelihood distributions of physical quantities derived from our fit to the observed SED: fraction of the total IR luminosity contributed by dust in the diffuse ISM ($f_\mu$); total 
effective V-band absorption optical depth of dust ($\hat\tau_V$); fraction of the total V-band absorption optical depth of the dust contributed by the diffuse ISM ($\mu$); 
effective V-band absorption optical depth of dust in the ambient ISM ($\mu\hat\tau_V$); specific star formation rate ($\psi_S$); stellar mass ($M_\ast$) and dust mass ($M_d$); total dust IR luminosity ($L_d^{\,\mathrm{tot}}$); global contributions by PAHs ($\xi_\mathrm{PAH}^{\,\mathrm{tot}}$), the hot mid-IR continuum ($\xi_\mathrm{MIR}^{\,\mathrm{tot}}$) and warm dust to $L_d^{\,\mathrm{tot}}$($\xi_\mathrm{W}^{\,\mathrm{tot}}$); 
temperature of warm dust in stellar birth clouds ($T_\mathrm{W}^{\,\mathrm{BC}}$); contribution by cold dust to \ldust\ ($\xi_\mathrm{C}^{\,\mathrm{tot}}$), 
and temperature of cold dust in the diffuse ISM ($T_\mathrm{C}^{\,\mathrm{ISM}}$).}
\label{fig:seds}
\end{figure*}

\setcounter{figure}{2}

\begin{figure*}
\begin{minipage} [t] {0.5\linewidth}
\centering
\includegraphics[width=\textwidth]{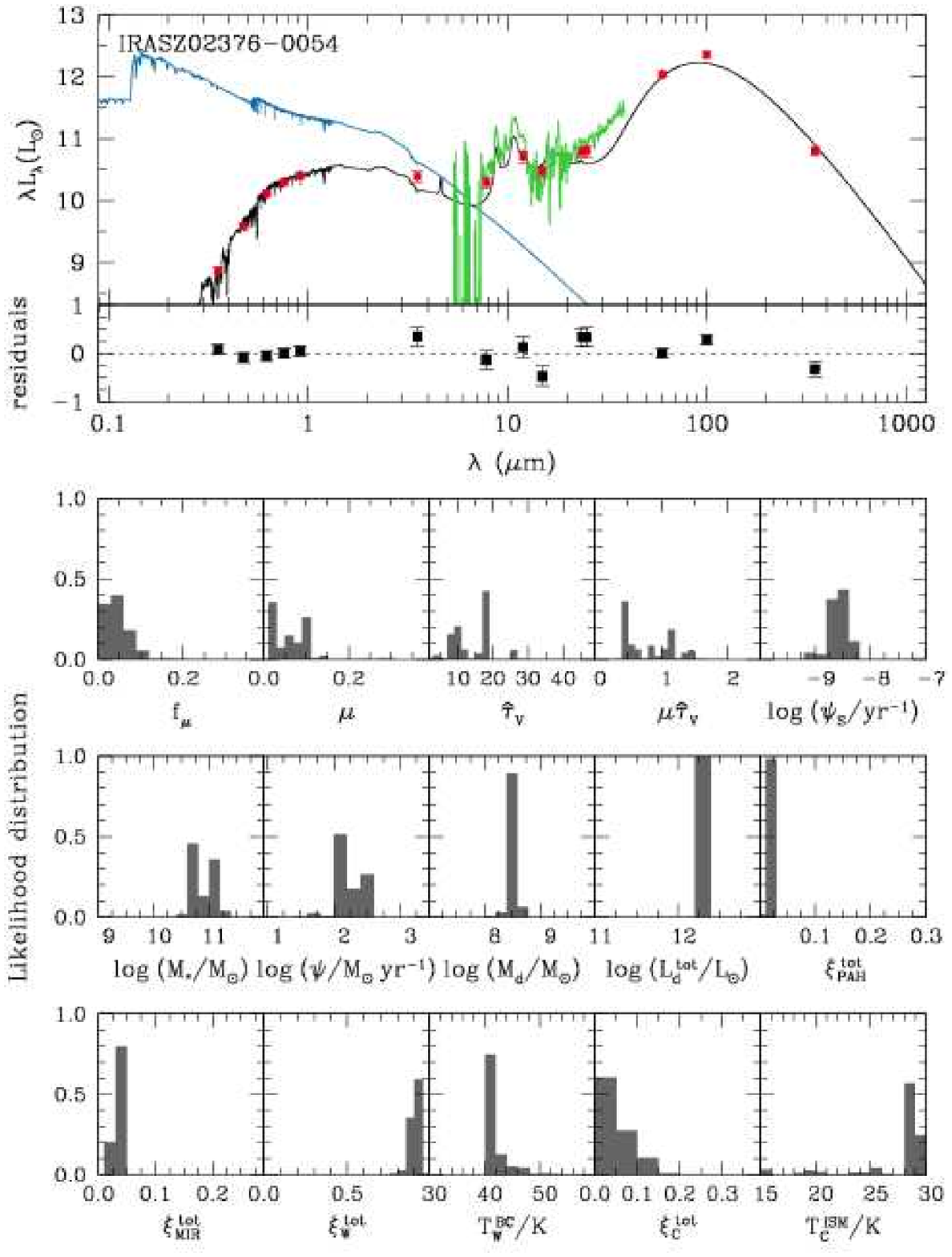}
\end{minipage}
\begin{minipage}[t] {0.5\linewidth}
\centering
\includegraphics[width=\textwidth]{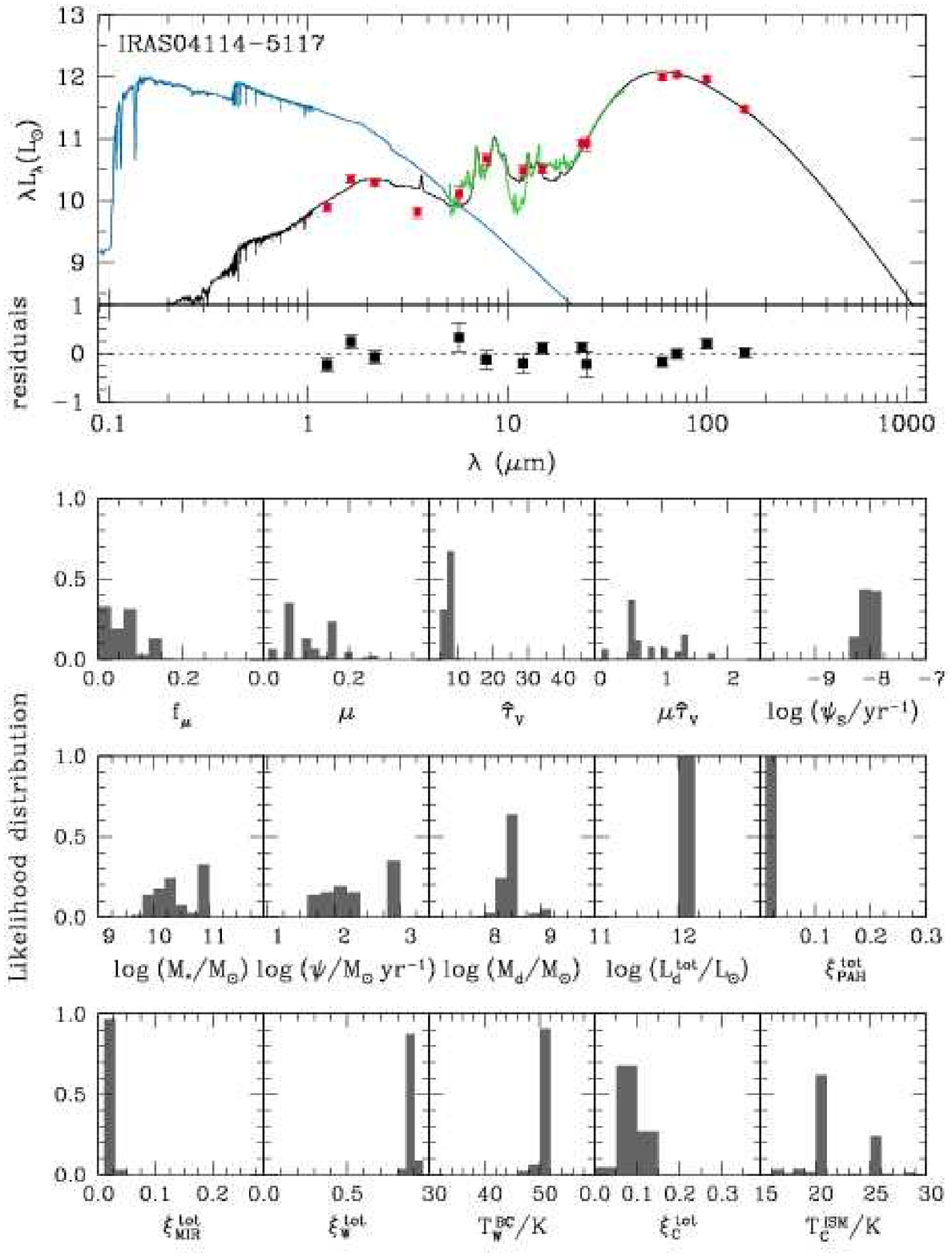}
\end{minipage}
\begin{minipage} [b] {0.5\linewidth}
\centering
\includegraphics[width=\textwidth]{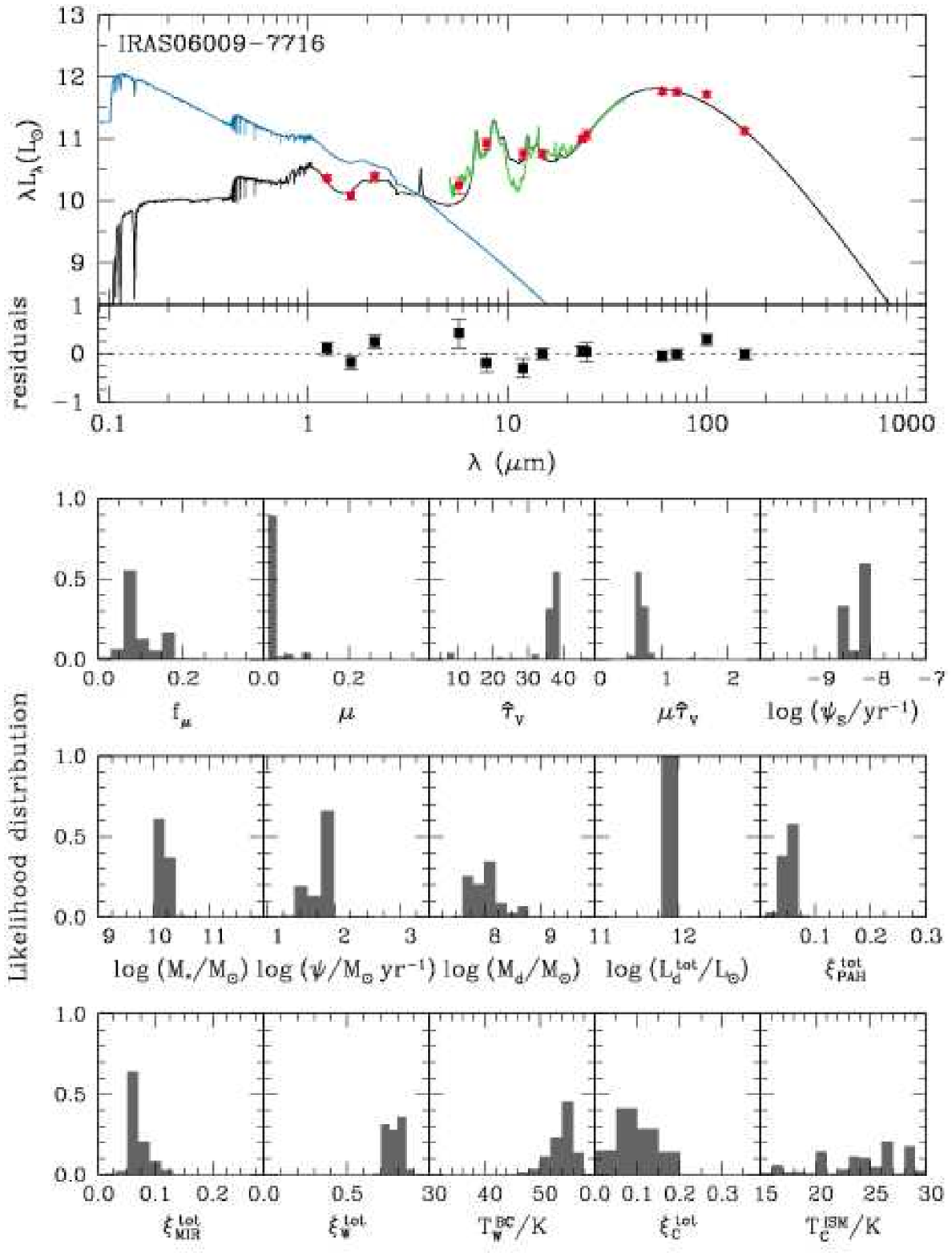}
\end{minipage}
\begin{minipage}[b] {0.5\linewidth}
\centering
\includegraphics[width=\textwidth]{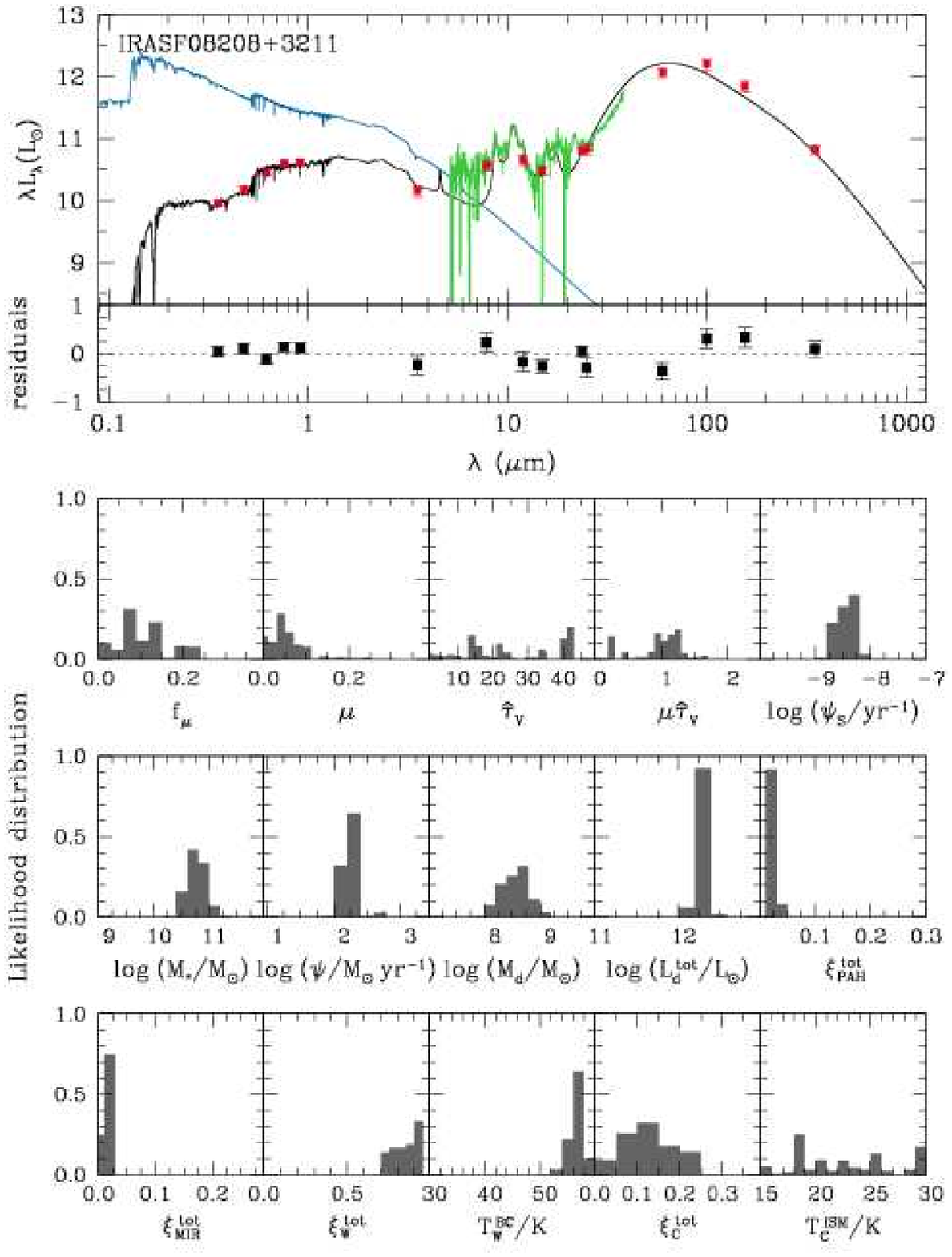}
\end{minipage}
\caption{{\it Continued.}}
\end{figure*}

\setcounter{figure}{2}
\begin{figure*}
\begin{minipage} [t] {0.5\linewidth}
\centering
\includegraphics[width=\textwidth]{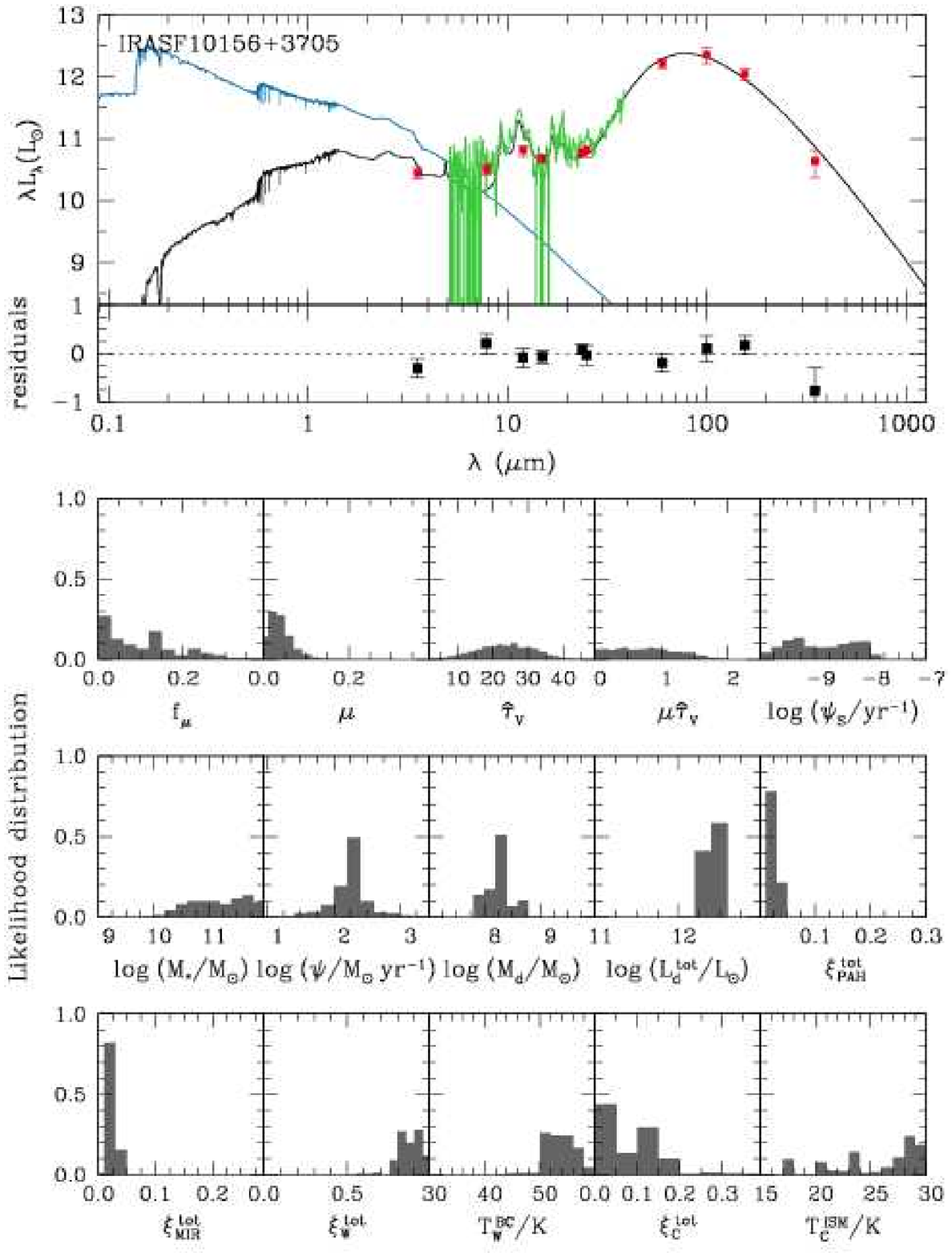}
\end{minipage}
\begin{minipage}[t] {0.5\linewidth}
\centering
\includegraphics[width=\textwidth]{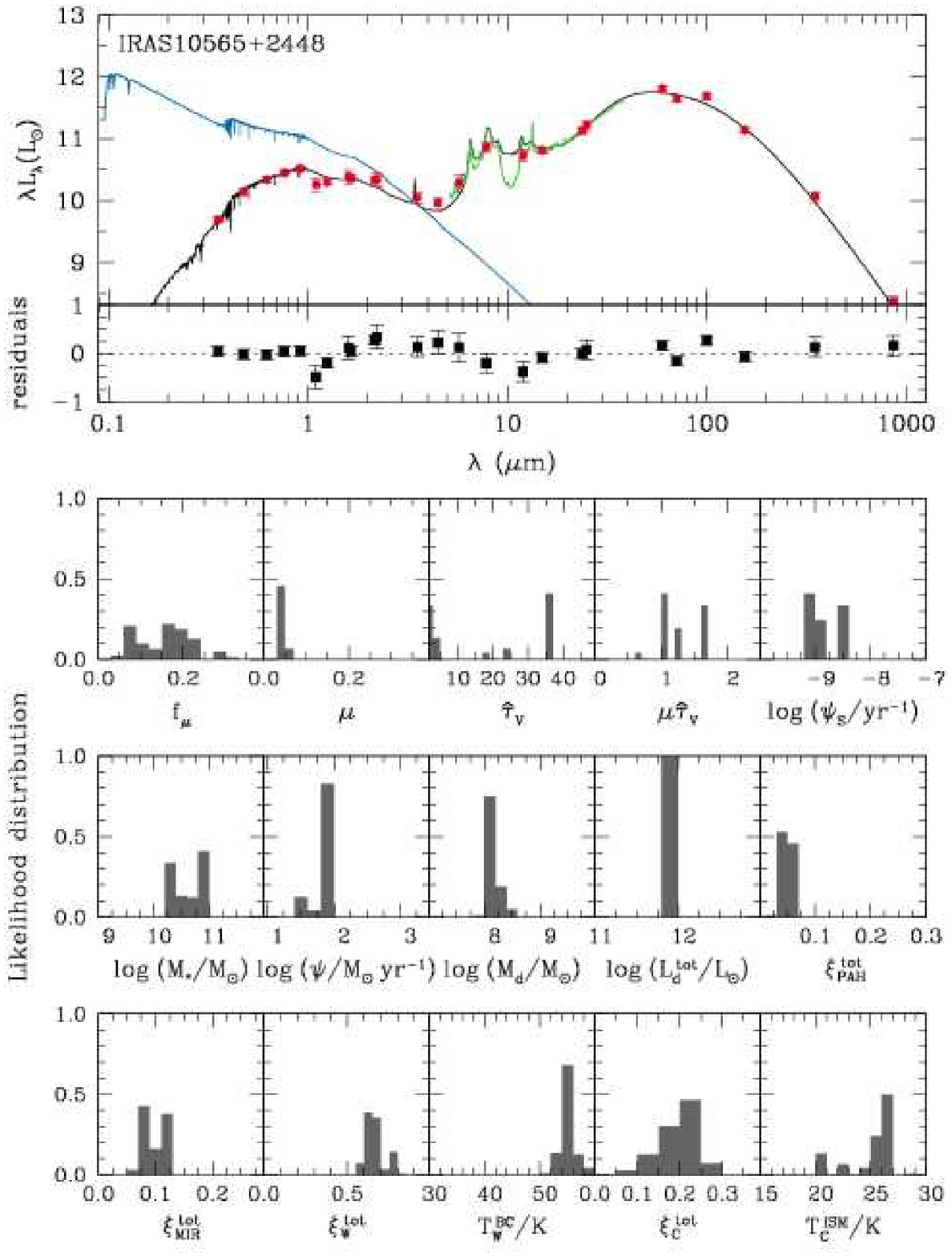}
\end{minipage}
\begin{minipage} [b] {0.5\linewidth}
\centering
\includegraphics[width=\textwidth]{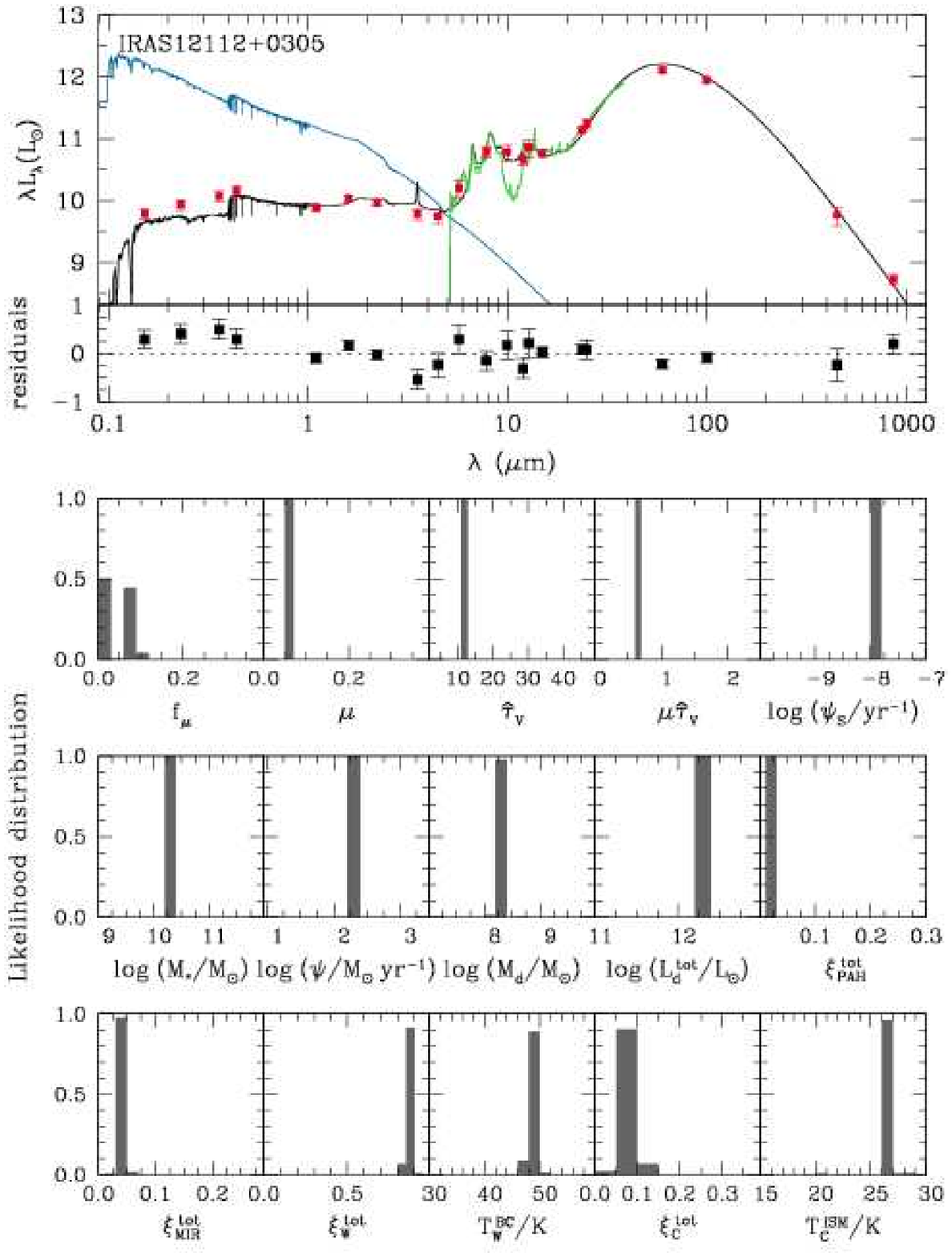}
\end{minipage}
\begin{minipage}[b] {0.5\linewidth}
\centering
\includegraphics[width=\textwidth]{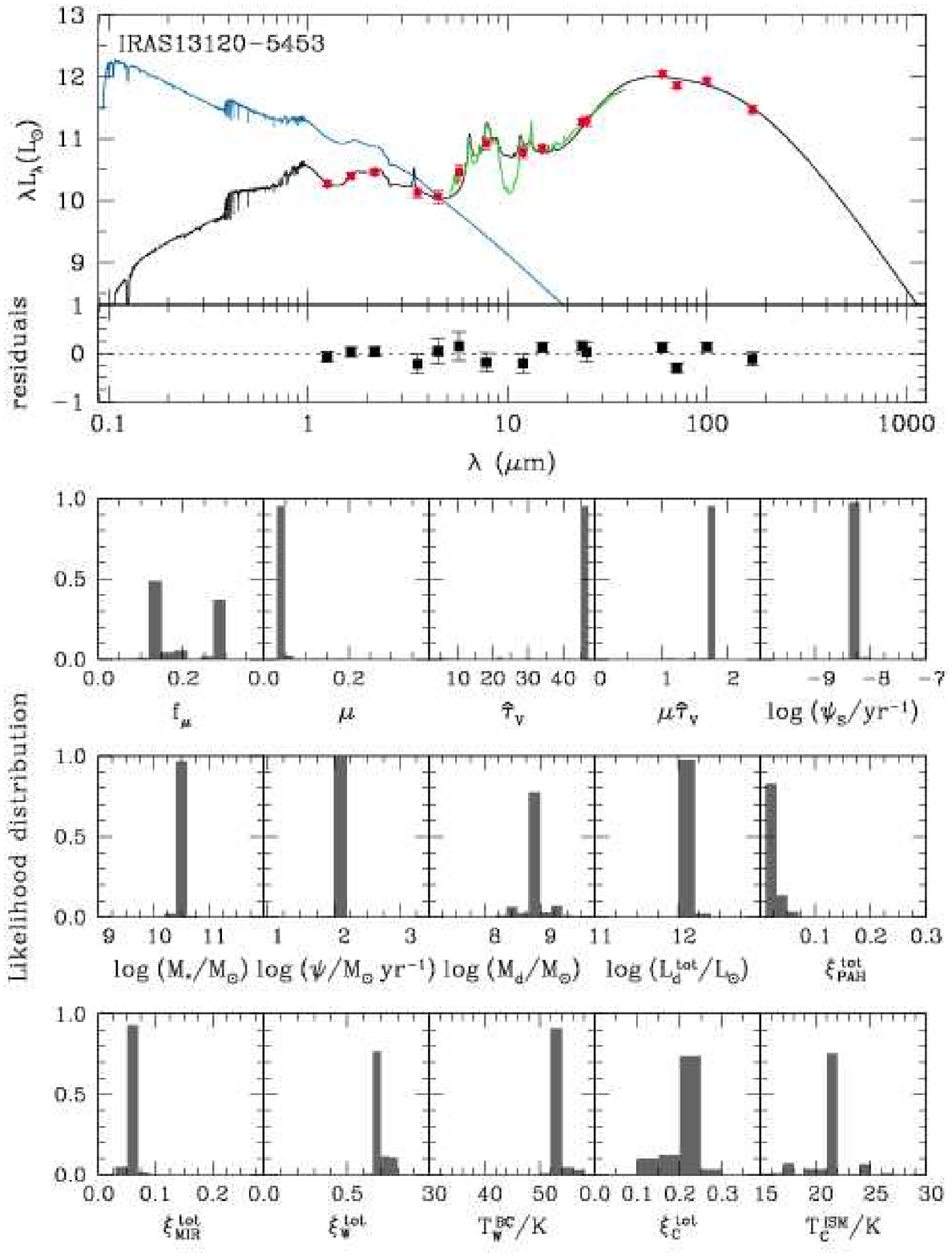}
\end{minipage}
\caption{{\it Continued.}}
\end{figure*}

\setcounter{figure}{2}
\begin{figure*}
\begin{minipage} [t] {0.5\linewidth}
\centering
\includegraphics[width=\textwidth]{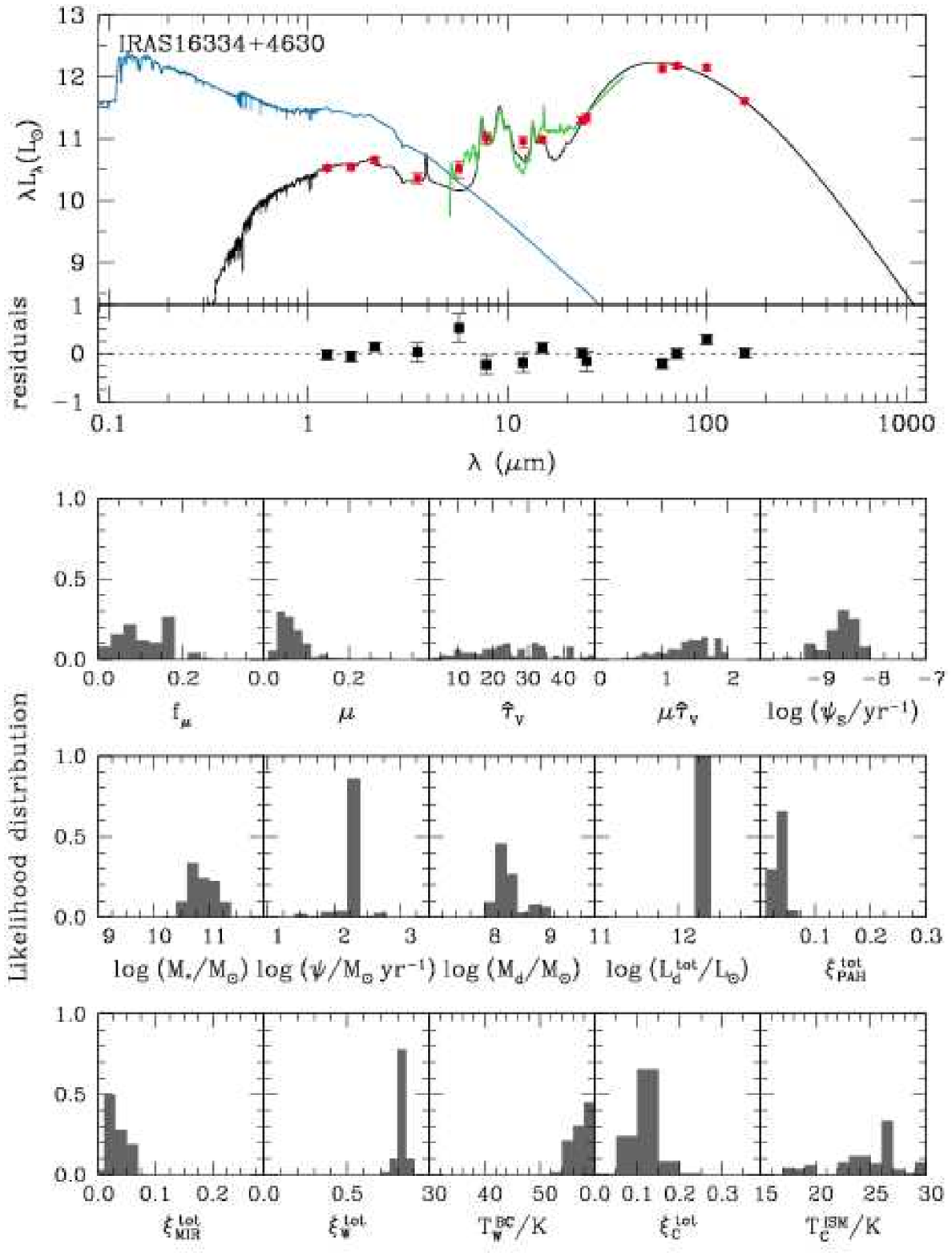}
\end{minipage}
\begin{minipage}[t] {0.5\linewidth}
\centering
\includegraphics[width=\textwidth]{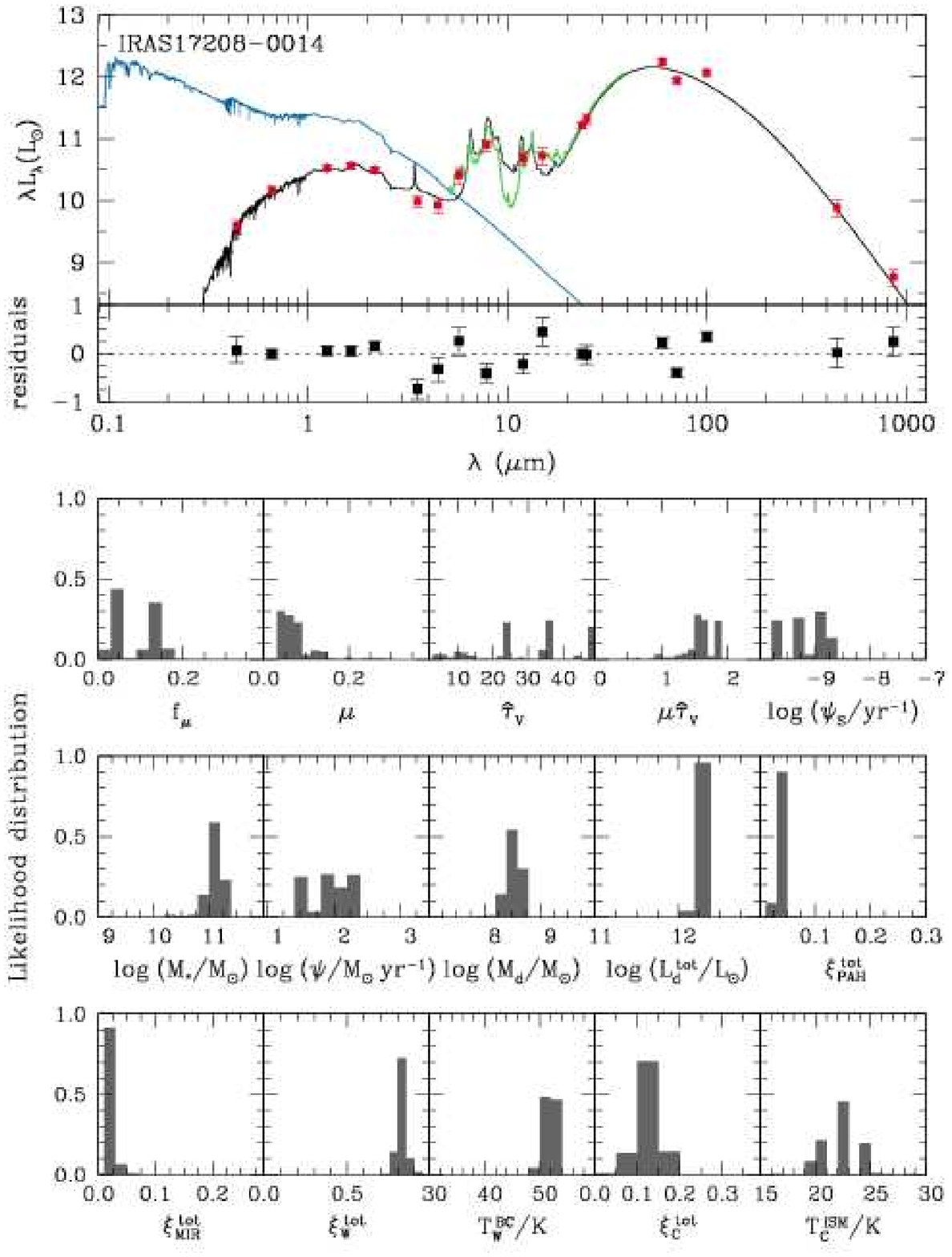}
\end{minipage}
\begin{minipage} [b] {0.5\linewidth}
\centering
\includegraphics[width=\textwidth]{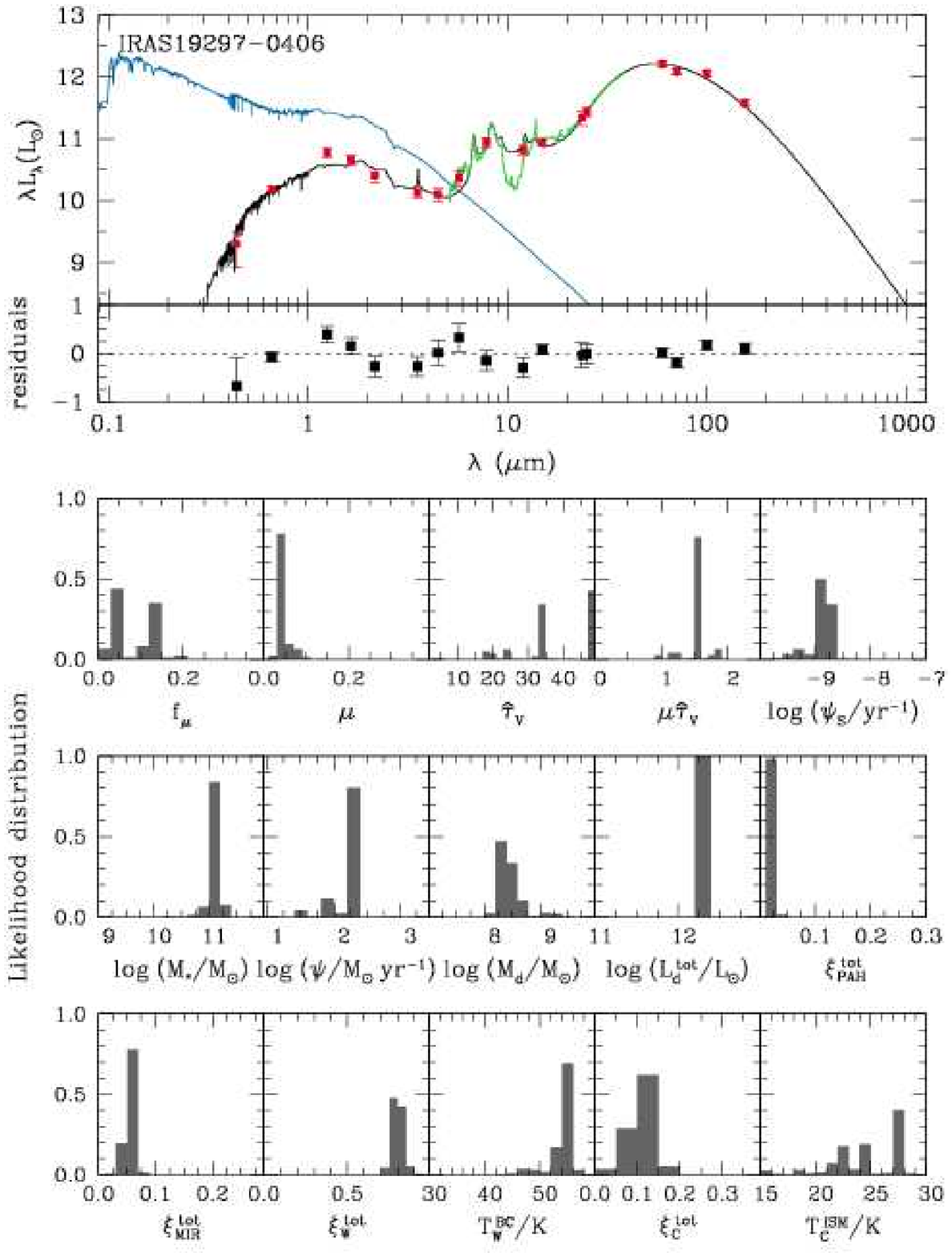}
\end{minipage}
\begin{minipage}[b] {0.5\linewidth}
\centering
\includegraphics[width=\textwidth]{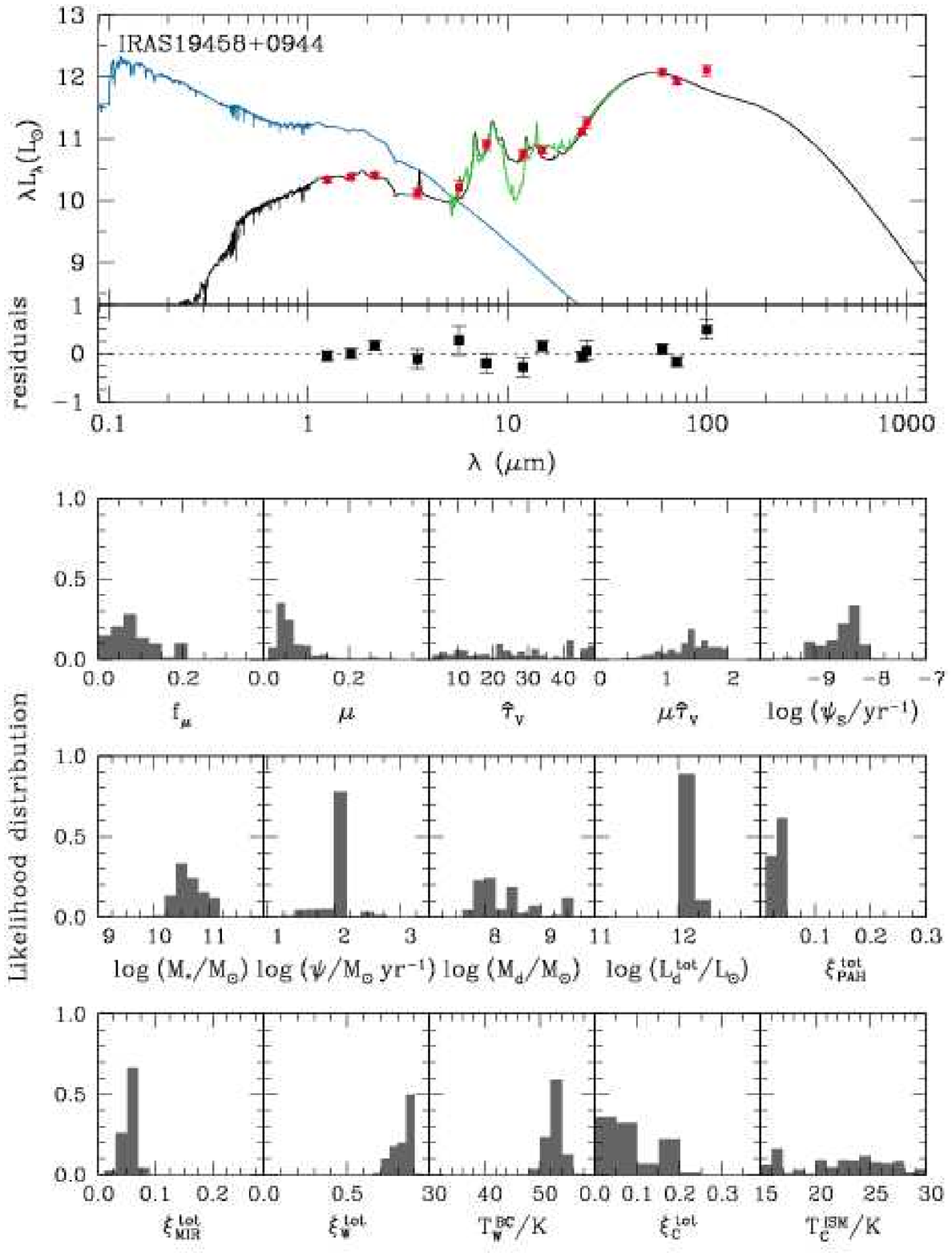}
\end{minipage}
\caption{{\it Continued.}}
\end{figure*}

\setcounter{figure}{2}
\begin{figure*}
\begin{minipage} [t] {0.5\linewidth}
\centering
\includegraphics[width=\textwidth]{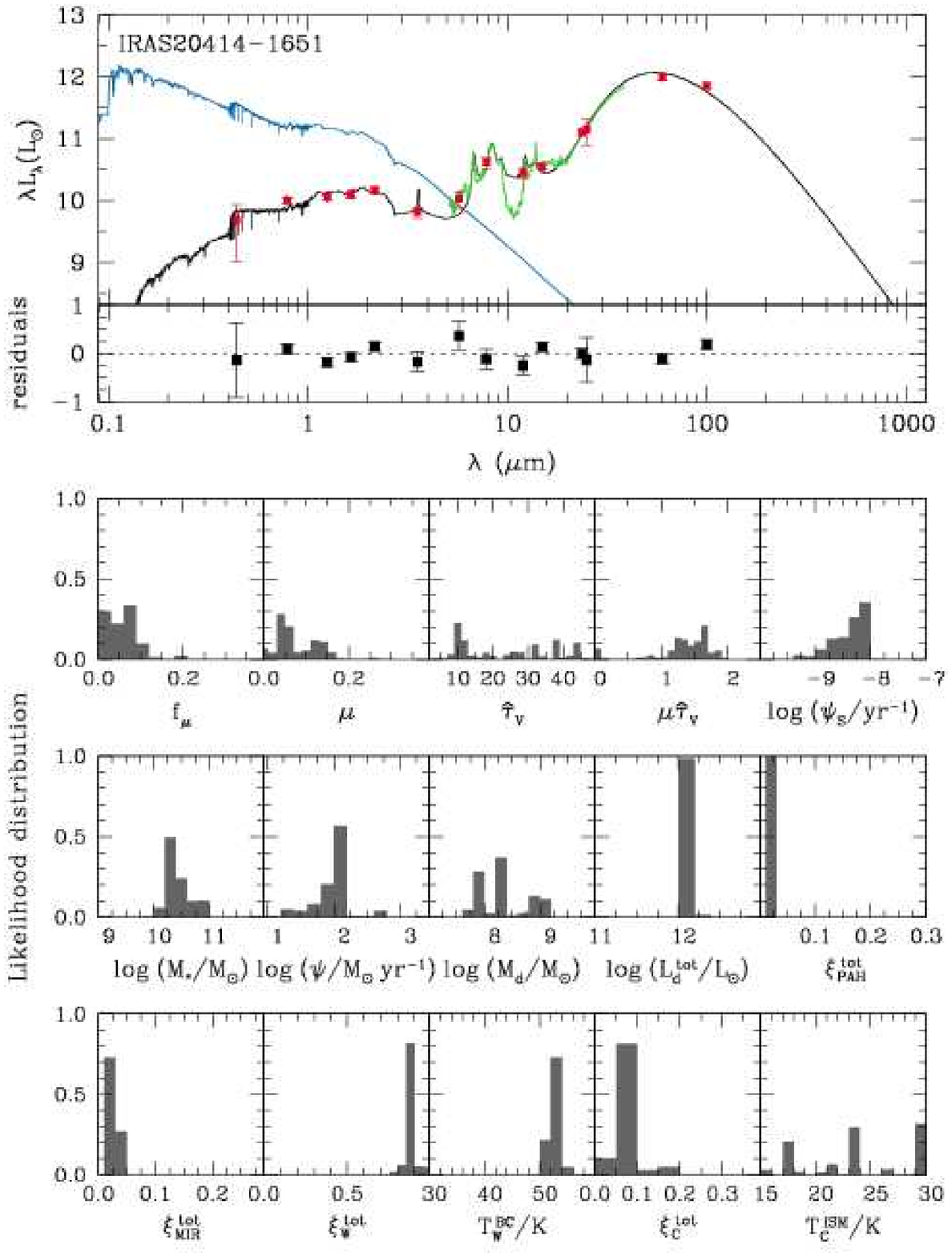}
\end{minipage}
\begin{minipage}[t] {0.5\linewidth}
\centering
\includegraphics[width=\textwidth]{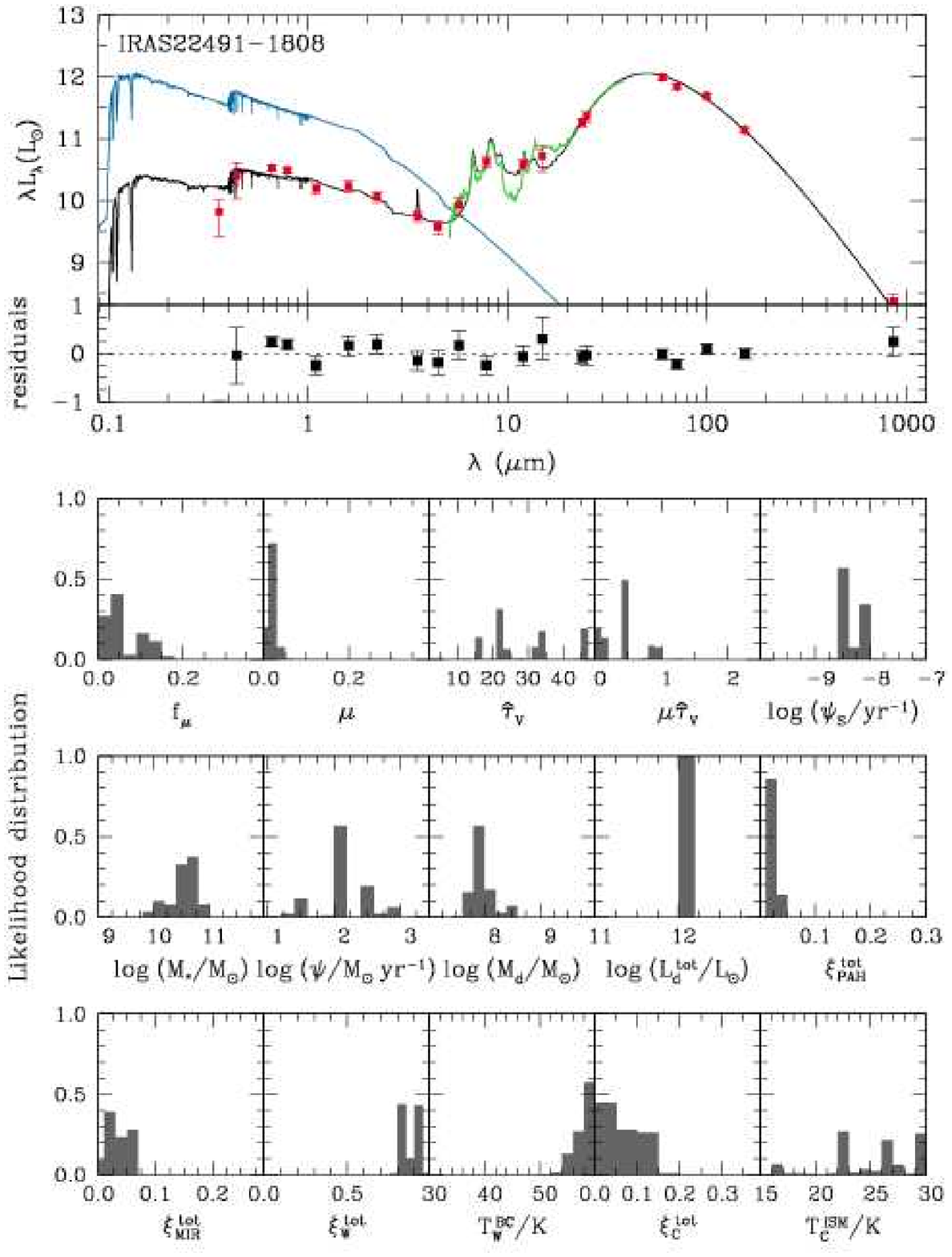}
\end{minipage}
\caption{{\it Continued.}}
\end{figure*}

\begin{figure}
\begin{center}
\includegraphics[width=0.4\textwidth]{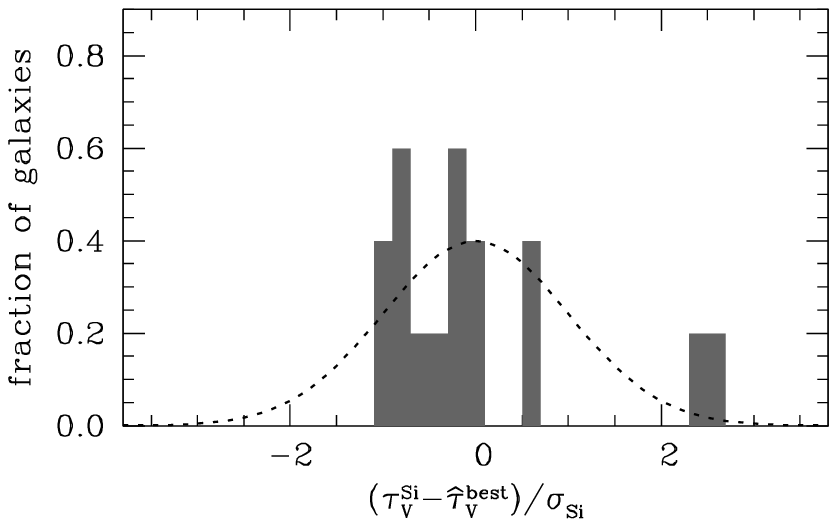}
\caption{Distribution of the difference between the $V$-band optical depth inferred from the observed silicate strength $\tau_V^\mathrm{Si}$
and the best-fit model $V$-band optical depth of the birth cloud component $\hat\tau_V^\mathrm{best}$, in units of the measurement error $\sigma_\mathrm{Si}$.
We assume a 30\% uncertainty to account for measurement errors from the IRS spectra and also for uncertainties in converting
from silicate strength to $V$-band optical depth.}
\label{fig:tau}
\end{center}
\end{figure}

In Fig.~\ref{fig:seds}, we show the results of our SED fitting for each of our 16 ULIRGs. We compare, in the top panels, the best-fit models (in black) to the observed multi-wavelength SEDs (in red). The residuals are shown in the bottom of each spectrum. The blue line shows the unattenuated emission from stars. For all our galaxies, this emission is very strong and blue, indicating that the stellar component is dominated by young stellar populations resulting from current or very recent star formation. Most of this stellar radiation is absorbed by dust and re-emitted in the infrared, leading to the large observed infrared-to-optical ratios in these galaxies. The infrared emission of our galaxies peaks typically at wavelengths shorter than 100~\mic, indicating overall high dust temperatures. For each galaxy, as a reference, we also plot the observed {\it Spitzer}/IRS spectrum (green line). We note that, even if we did not intend to fit the detailed shape of the mid-IR emission, our best-fit SED and the IRS spectrum are in good agreement in most cases. Both the shape of the PAH emission and the slope of the hot mid-infrared continuum are reproduced remarkably well considering the simplicity of the model. However, the fits do not reproduce the observed silicate absorption feature because dust self-absorption is not included in our infrared SEDs. This should have a negligible contribution in terms of the energy balance performed here.

Even if we do not explicitly include the silicate absorption in our infrared SEDs, the silicate optical depth measured from the observed IRS spectra provides valuable information about the optical depth, as discussed in Section~\ref{method}.
We find that in heavily dust-enshrouded systems as our ULIRGs, constraining well the dust optical depth is fundamental in our SED fitting and has a high impact in the statistical constraints of star formation rates, stellar masses, and contribution by the diffuse ISM and the birth clouds components to the dust heating. To illustrate how well the $V$-band optical depth in the birth clouds is recovered by our best-fit model, in Fig.~\ref{fig:tau}, we plot the distribution of the difference between the observational ($\tau_{V}^\mathrm{Si}$) and the best-fit (\tauv) $V$-band optical depth for the galaxies in our sample. In most cases, the $V$-band optical depth is recovered within $2\sigma$ (i.e. 60\% of $\tau_V^\mathrm{Si}$). This is a satisfactory result considering the uncertainties in deriving $\tau_V^\mathrm{Si}$ from the observed IRS spectra. 

An advantage of our method is that it enables us not only to find a best fit model, but also to compute the full likelihood distribution of each model parameter
and identify possible degeneracies between parameters. This allows us to determine how well the parameters are constrained and how the available set of observations affects these constraints (see more details in section~3.2 of \citealt{daCunha2008}). The bottom panels of Fig.~\ref{fig:seds} show the likelihood distribution of several physical properties of our ULIRGs derived from the comparison of the observed SEDs with our stochastic library of models as described in Section~\ref{fits}. We list the median-likelihood estimates and confidence intervals of some important parameters for each galaxy in Table~\ref{tab:results}.
The likelihood distributions of Fig.~\ref{fig:seds} show that all the parameters are reasonably well constrained for most of our galaxies. However, we can see some differences in how well some parameters are constrained from galaxy to galaxy, and this is primarily due to the different set of available photometry for each galaxy. Clearly, the more available data points, the better constrained are the parameters.

It is important to note how essential it is to include ultraviolet to near-infrared observations, even if the emission from ULIRGs is largely dominated by longer wavelengths. UV to near-IR observations probe stellar populations older than $10^8$~yrs which heat the diffuse ISM (i.e. stellar population which do not belong to the central, heavily obscured starburst). These observations help constrain the relative amount of dust luminosity heated by these stars (\fmu), the dust attenuation in the diffuse ISM ($\mu\tauv$), the stellar mass (\mstar) and, as a consequence, the specific star formation rate (\ssfr). This is visible in the case of IRASF10156+3705, for which no observations short-ward of 3.6~\mic\ are available, leading to typically more extended likelihood distributions (and hence wider confidence ranges) for these parameters. We note, however, that even in this case, the star formation rate is still relatively well-constrained because, given the input optical depth derived from the IRS spectrum, the model determines what is the necessary amount of star formation to power the very large infrared luminosity of the galaxy (constrained by the infrared observations).

In the infrared, given the wealth of available data at least up to 160~\mic, the total dust luminosity (\ldust) and the contribution to the total emission by PAHs (\xipahstot), the hot mid-IR continuum (\ximirtot), and warm dust in thermal equilibrium (\xiwarmtot), as well as the temperature of this dust (\tbgswarm), are well constrained in most of the cases.
Cold dust emits in the far-infrared to sub-millimetre, and hence is constrained by the available observations from 160 to 850~\mic. In general, the contribution of this component to the total emission (\xicoldtot) is well constrained by these observations. However, the equilibrium temperature of the cold dust component in our model (\tbgscold) is relatively hard to constrain for our ULIRGs even with far-infrared data, because the contribution of this dust to the total infrared emission is small and hard to isolate from the dominating warm dust component. We note that this contributes to the broadening of the likelihood distributions of the total dust mass, since cold dust, even while contributing little to the total infrared luminosity, contributes significantly to the dust mass.

We have compared our median-likelihood estimates of the total dust luminosity \ldust\ with the infrared luminosities empirically derived using the \iras\ fluxes, $L_\mathrm{IR}$, presented in Table~\ref{tab:basic}. The two luminosities agree within typically 0.07~dex (with no systematic offset)
for most of the galaxies. For only four galaxies, the difference between the two luminosities is bigger:
IRASF08208+3211, IRASF10156+3705, IRAS17208-0014 and IRAS19458+0944, although even in these cases the maximum offset between the two luminosities is at most 0.14~dex (i.e.~38\%).

It is important to check how our stellar mass estimates compare with the results of previous studies, specially considering the very large optical depths
of ULIRGs. \cite{Rodriguez2010} derived the stellar masses of 36 local ULIRGs using optical spectroscopy and found that ULIRGs have moderate stellar masses $\lesssim 1.4\times10^{11}~\msun$ (using a \citealt{Salpeter1955} IMF and \citealt{Bruzual2003} models spectral synthesis models; see also, e.g.~\citealt{Genzel2001,Tacconi2002}). 
We note that the use of a Salpeter IMF and the \cite{Bruzual2003} models can make the \mstar\ estimates systematically higher by 0.15--0.3~dex compared to the use of a Chabrier IMF and the latest version of the Bruzual \& Charlot models used in our study \citep{Bruzual2007,daCunha2009}.
Nevertheless, we find that our estimates of the stellar masses of our ULIRGs are consistent with the previous result of \cite{Rodriguez2010}, except for IRASF10156+3705, for which we find a significantly larger \mstar, but this may be attributed to the more uncertain estimate of \mstar\ for this galaxy due to the lack of optical and near-infrared data (as mentioned above).

\section{Discussion}\label{discussion}

\subsection{Comparison between star formation rate and infrared luminosity}

To test the validity of our approach, we compare our star formation rate estimates with the ones obtained from the infrared luminosity using the formula of \cite{Kennicutt1998}.
This formula assumes that all the infrared emission from the galaxy is powered by recent star formation, i.e. all the energy emitted by young stars in the galaxy is absorbed and re-radiated by dust, and no dust is heated by old stellar populations ({\it optically-thick starburst}).
In our model, this corresponds to having very small values of the \fmu\ parameter, i.e. the contribution by the diffuse ISM (heated by stars older than 100~Myr) to the total dust luminosity \ldust\ is very small (see Table~\ref{tab:results}). Also, to power the huge infrared luminosities of ULIRGs with young stars alone, the effective attenuation in the birth clouds must be very high, as well as the star formation rates. Therefore we expect the star formation rates of our ULIRGs to be similar to the ones derived using the Kennicutt formula.

\begin{figure}
\begin{center}
\includegraphics[width=0.45\textwidth]{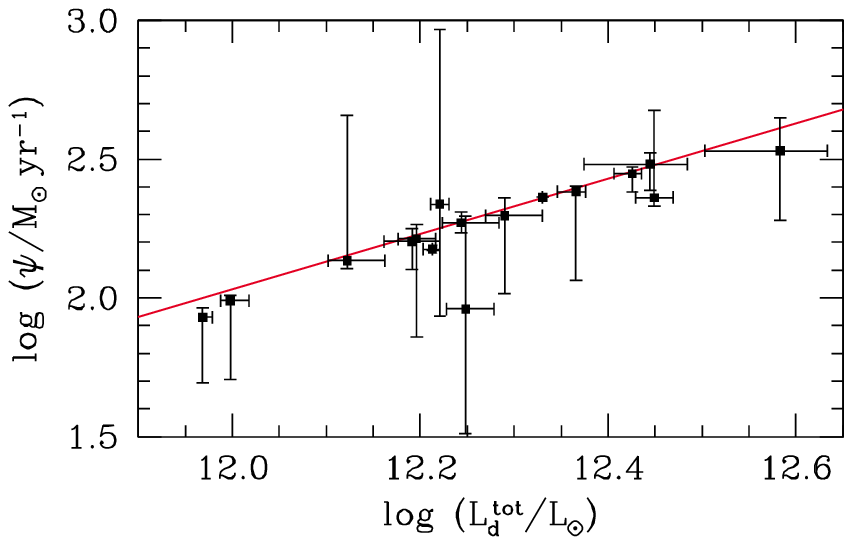}
\caption{Median-likelihood estimate of the total dust luminosity plotted against star formation rate for the ULIRGs in our sample (both derived from our SED fitting). The red line shows the Kennicutt (1998) relation for optically-thick starbursts, corrected for Chabrier (2003) IMF (eq.~\ref{sfr_k98}). The error bars correspond to the 16th -- 84th percentile range of the likelihood distributions of each parameter.}
\label{fig:sfr_k98}
\end{center}
\end{figure}

\begin{table}
\caption{Comparison between the median-likelihood estimates of the star formation rate derived from our fits to the SEDs (a) and the star formation rate computed from the dust luminosity using the formula of \citealt{Kennicutt1998} (b).} \label{tab:sfr}
\begin{center}
\begin{tabular}{lccccccccccc} \hline \hline
 & &  \\
Galaxy & $\psi/M_{\odot}$~yr$^{-1}$ & $\psi_{\mathrm{K98}}/M_{\odot}$~yr$^{-1}$\\ 
 & (a) & (b) \\
 \hline
 \\[-0.75ex]
IRAS00199-7426  &  $198^{+29}_{-86}$  &  $210$ \\[0.75ex]
IRAS01494-1845  &  $186^{+16}_{-13}$  &  $189$ \\[0.75ex]
IRASZ02376-0054  &  $230^{+174}_{-11}$  &  $304$ \\[0.75ex]
IRAS04114-5117  &  $218^{+632}_{-117}$  &  $180$ \\[0.75ex]
IRAS06009-7716  &  $98^{+4.1}_{-42}$  &  $108$ \\[0.75ex]
IRASF08208+3211  &  $303^{+21}_{-43}$  &  $301$ \\[0.75ex] 
IRASF10156+3705  &  $339^{+72}_{-100}$  &  $414$ \\[0.75ex]
IRAS10565+2448  &  $85^{+6.9}_{-34}$  &  $100$ \\[0.75ex]
IRAS12112+0305  &  $230^{+0.0}_{-0.0}$  &  $231$ \\[0.75ex]
IRAS13120-5453  &  $150^{+0.0}_{-1.7}$  &  $176$ \\[0.75ex]
IRAS16334+4630  &  $280^{+14}_{-33}$  &  $288$ \\[0.75ex]
IRAS17208-0014  &  $91^{+102}_{-56}$  &  $191$ \\[0.75ex]
IRAS19297-0406  &  $241^{+10}_{-116}$  &  $251$ \\[0.75ex]
IRAS19458+0944  &  $160^{+16}_{-30}$  &  $168$ \\[0.75ex] 
IRAS20414-1651  &  $163^{+18}_{-84}$  &  $170$ \\[0.75ex]
IRAS22491-1808  &  $136^{+297}_{-8.4}$  &  $143$ \\[0.75ex]
\hline
\end{tabular}
\end{center}
\end{table}

In Fig.~\ref{fig:sfr_k98} and Table~\ref{tab:sfr}, we compare our median-likelihood estimates of the star formation rates of our ULIRGs with the star formation rates predicted from the total infrared luminosity using the \cite{Kennicutt1998} formula for optically-thick starbursts (corrected for a Chabrier IMF). The star formation rate is computed as:
\begin{equation}
\frac{\psi_\mathrm{K98}}{M_{\odot}\,\mathrm{yr}^{-1}}=1.08\times10^{-10}~\frac{L_\mathrm{IR}}{L_{\odot}}\,,
\label{sfr_k98}
\end{equation}
where $L_\mathrm{IR}$ is the total infrared luminosity between 8 and 1000~\mic, which we approximate with the total infrared luminosity derived from our fits, \ldust.

We verify that our star formation rates agree remarkably well with the ones predicted using the infrared luminosities. The difference between \sfr\ and 
$\psi_\mathrm{K98}$ is typically less than $20\%$ (except for IRASZ02376-0054 and IRAS17208-0014, but even in those cases $\psi_\mathrm{K98}$ is still
in the confidence range of \sfr) . This confirms that the model is correctly attributing most of the infrared emission of the ULIRGs to the starburst phase (i.e. the `birth cloud' component in our model). This was only achieved through a correct estimation of the optical depths enabled by the use of the IRS spectra.

We caution that in more quiescent galaxies, dust heated by old stars in the ambient ISM contributes more significantly to the total infrared luminosity, and the star formation rates obtained using \cite{daCunha2008} may differ significantly than those derived using the formula of \cite{Kennicutt1998}. In such cases the star formation rates would be overestimated using the Kennicutt law.

\subsection{The star formation histories of ULIRGs}\label{sfh}

\begin{figure}
\begin{center}
\includegraphics[width=0.45\textwidth]{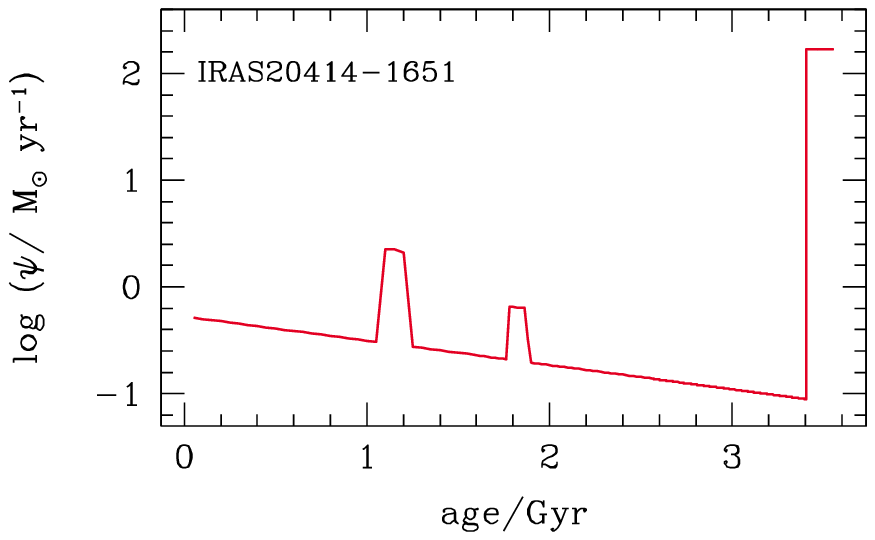}
\caption{Example of best-fit model star formation history of one of our ULIRGs, IRAS20414-1651, with a current age $3.56$~Gyr.}
\label{fig:sfh}
\end{center}
\end{figure}

The model described in Section~\ref{sed_mod} allows us to explore possible star formation histories of galaxies, and the effect they have on observable properties. One interesting result of the SED fitting of our ULIRGs is that, for all 16 of them, the star formation history corresponding to the best-fit model (Fig.~\ref{fig:seds}) shows a strong burst of star formation occurring at the present age, i.e. all the best-fit models are clearly ones with a current starburst (see example of Fig.~\ref{fig:sfh}).
We find that, on average, this burst of star formation is responsible for forming $30\%$ of the measured stellar mass of our ULIRGs over the last 100~Myr (this corresponds to about $2\times10^{10}$~\msun\ formed during the burst).

The observed spectral energy distributions of our ULIRGs can only be well fitted with a simple, exponentially-declining star formation history with superimposed random bursts (as we use in our stochastic library of models; see Section~\ref{library}) if the star formation history contains a strong burst of star formation at the present age.
This star formation history, while not necessarily being a unique solution, is a plausible scenario consistent with the main result that can be inferred from the SED fitting -- that the galaxy must be experiencing a strong burst of star formation where a significant fraction of its stellar mass is being formed.
To constrain the detailed star formation history of a galaxy, we would require optical spectroscopic data to characterize the stellar populations of different ages (e.g.~\citealt{Rodriguez2008,Rodriguez2010}). Nevertheless, our results show that fitting the broad-band SEDs can at least provide some clues on the star formation mode of these galaxies. In particular, our fits clearly show that, in order to
reproduce the observed properties of our ULIRGs, a star formation history with a strong current burst of star formation is required.

\subsection{Comparison with local galaxies of infrared luminosities between $10^{10}$ and $10^{12}~\lsun$}

\begin{figure*}
\begin{center}
\includegraphics[width=0.75\textwidth]{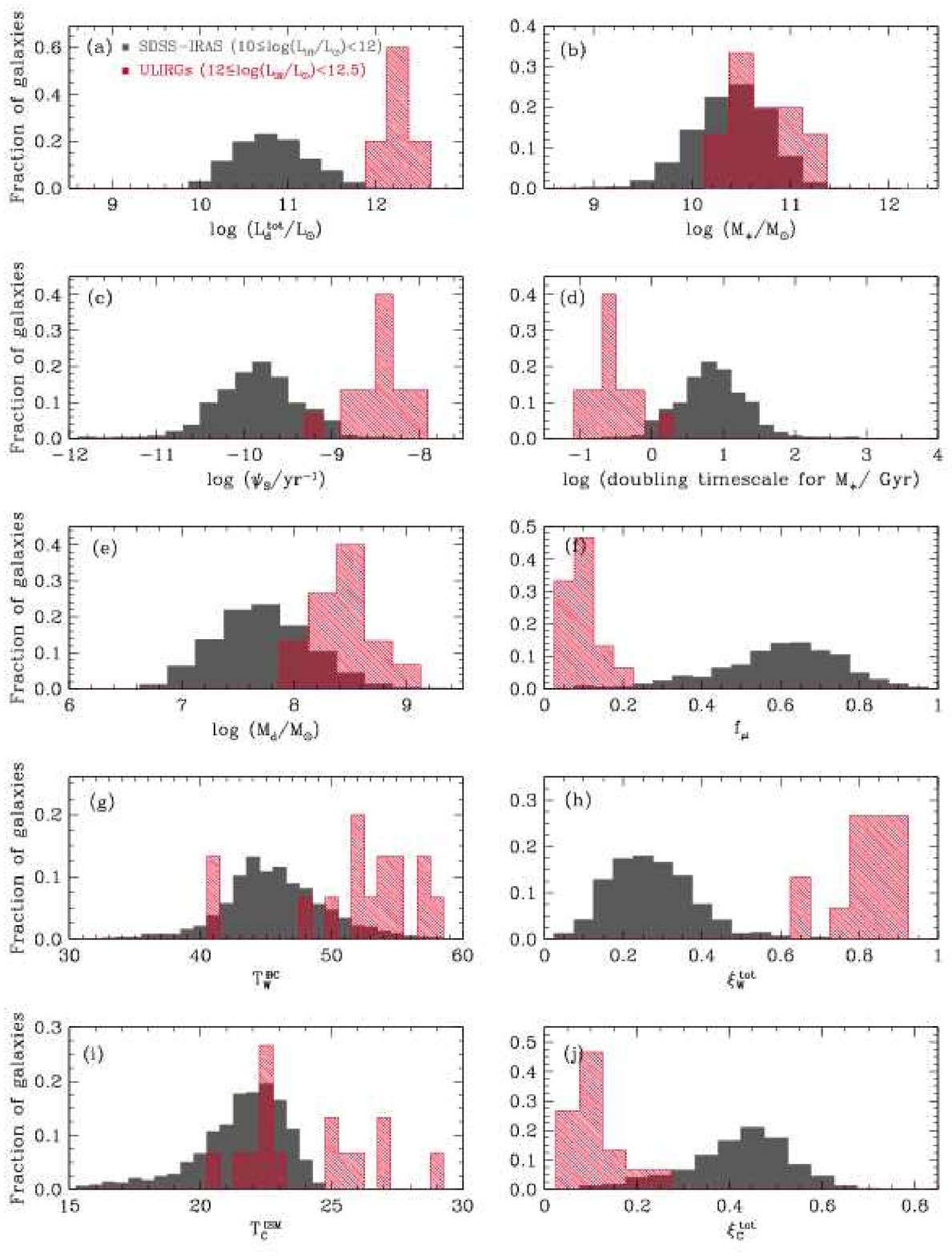}
\caption{Comparison between the distributions of median-likelihood estimates of the physical parameters of 15 of our ULIRGs with $12 \le \log(\ldust/\lsun) < 12.5$ (red histograms), and of 1538 local star-forming galaxies of lower infrared luminosities ($10 \le \log(\ldust/\lsun) < 12$) selected from the SDSS-\iras\ sample of \citealt{daCunha2009} (grey histograms).
(a) dust luminosity, \ldust; (b) stellar mass, \mstar; (c) Specific star formation rate averaged over the last 100~Myr, \ssfr; (d) doubling timescale for the stellar mass, assuming the galaxies would continue forming stars at the present rate; (e) dust mass, \mdust; (f) fraction of total dust luminosity contributed by the diffuse ISM, \fmu; (g) temperature of warm dust in the birth cloud component, \tbgswarm; (h) fraction of total dust luminosity contributed by warm dust in thermal equilibrium, \xiwarmtot; (i) temperature of cold dust in the diffuse ISM, \tbgscold; (j) fraction of total dust luminosity contributed by cold dust in thermal equilibrium, \xicoldtot.}
\label{fig:hist}
\end{center}
\end{figure*}

One of the main advantages of the method used in this paper is that our results are directly comparable to results found for other samples of galaxies using the same framework.
The physical properties of a sample of 3258 local star-forming SDSS galaxies ($z \lesssim 0.1$) detected with \iras, for which \galex\ and 2MASS observations are also available (SDSS-\iras\ sample from now on), were derived in \cite{daCunha2009} using the technique used this paper.
In this section, we compare the physical properties of local galaxies in two bins of infrared luminosity: $10 \le \log(L_\mathrm{IR}/\lsun) < 12$ (1538 galaxies with high S/N observations selected from the SDSS-\iras\ sample of \citealt{daCunha2009}), and $12 \le \log(L_\mathrm{IR}/\lsun) < 12.5$ (15 ULIRGs from the present study). This is done in order to examine whether our findings based on SED modelling are consistent with the current understanding of ULIRGs.
In Fig.~\ref{fig:hist}, we plot the distributions of the median-likelihood estimates of several physical parameters related to the star formation activity and the dust properties of the galaxies. We describe the main similarities and differences between the two samples, even considering the low number of ULIRGs compared to the number of lower-luminosity galaxies.

Fig.~\ref{fig:hist}(b) shows that the stellar masses of galaxies in the two infrared luminosity bins are similar, with the ULIRGs tending to be slightly more massive.
We note, however, that the stellar mass estimates of ULIRGs are typically more uncertain than those of galaxies of lower infrared luminosities (and lower dust attenuation).
In Fig.~\ref{fig:hist}(c) we show that ULIRGs have typically much higher specific star formation rates than the SDSS-\iras\ galaxies, implying doubling timescales of their stellar mass (assuming that they continue forming stars at the present rate) of the order of less than 1~Gyr [Fig.~\ref{fig:hist}(d)]. This is another confirmation of the well-known starbursting nature of our local ULIRGs (see, e.g.~\citealt{Genzel1998}).
 
We find that galaxies with $12 \le \log(L_\mathrm{IR}/\lsun) < 12.5$ also present very different properties of the dust emission than galaxies with $10 \le \log(L_\mathrm{IR}/\lsun) < 12$. Fig.~\ref{fig:hist}(f) shows that, in ULIRGs, the contribution of the diffuse ISM component to the total infrared luminosity, \fmu, is typically very small (about 0.10), compared with what is found for the SDSS-\iras\ galaxies. \cite{daCunha2009} find that, for these galaxies, dust heated by old stars in the diffuse ISM contributes significantly to the total infrared luminosity, with most values of \fmu\ typically between 0.40 and 0.70.

Furthermore, Figs.~\ref{fig:hist}(g) and (i) show that the typical equilibrium temperatures of dust grains in the `birth cloud' component, \tbgswarm, and in the `diffuse ISM' component, \tbgscold, are typically higher in ULIRGs than in the lower infrared luminosity galaxies. Additionally, the contribution of warm dust to the total infrared luminosity \xiwarmtot\ is very large for ULIRGs [Fig.~\ref{fig:hist}(h)], while the contribution of cold dust \xicoldtot\ is much smaller than in the other sample [Fig.~\ref{fig:hist}(j)]. Warm dust is the dominant contributor to the total dust emission in ULIRGs, and therefore, the overall temperature of dust in our star-forming ULIRGs is higher than that of the SDSS-\iras\ galaxies (see also \citealt{Clements2010}).

\begin{figure}
\begin{center}
\includegraphics[width=0.5\textwidth]{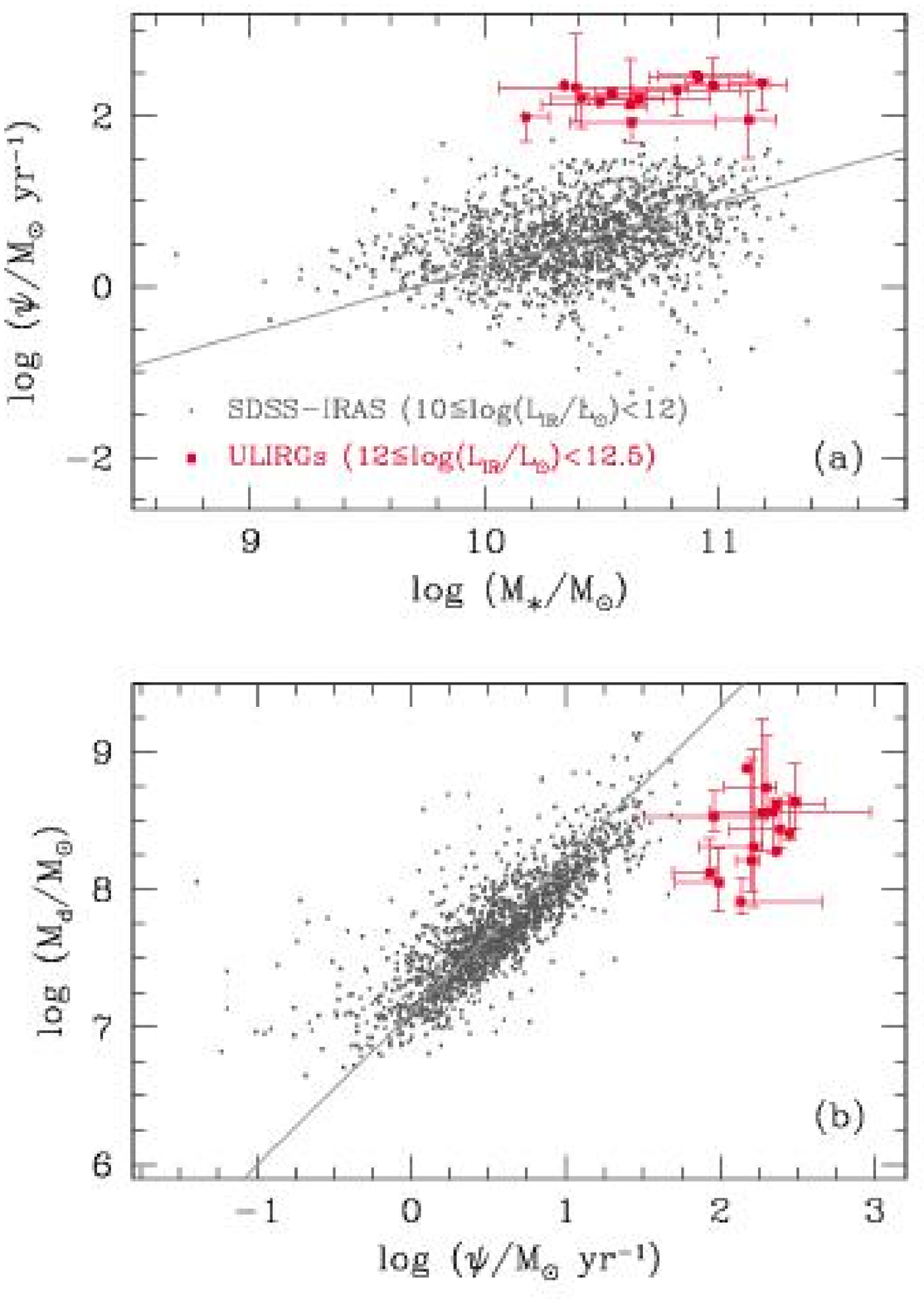}
\caption{Median-likelihood estimates of star formation rate averaged over the last $10^8$~yrs \sfr, plotted against stellar mass \mstar\ (a) and dust mass \mdust\ (b) for 15 of our ULIRGs with $12 \le \log(\ldust/\lsun) < 12.5$ (red points), and 1538 local star-forming galaxies of lower infrared luminosities ($10 \le \log(\ldust/\lsun) < 12$) selected from the SDSS-\iras\ sample of \citealt{daCunha2009} (grey points).
The errors bars correspond to the 16th--84th percentile range of the parameter likelihood distributions. The light grey lines show the fit to the SDSS points: (a) best fit using a slope of 0.77 as found in \cite{Elbaz2007}: $\log(\psi)=-7.467+0.77\log(\mstar)$; (b) $\log(\mdust)=7.107+1.11\log(\psi)$, as found by \cite{daCunha2009}.}
\label{fig:mdust}
\end{center}
\end{figure}

Based on the model fits we may also explore how the relations between the star formation rate, stellar mass and dust mass can be useful to study the star formation mode and evolution of our galaxies.
In Fig.~\ref{fig:mdust}(a), we plot our median-likelihood estimates of the star formation rate against the stellar mass for our 15 local ULIRGs with $12 \le \log(L_\mathrm{IR}/\lsun) < 12.5$, and the 1538 SDSS-\iras\ galaxies with $10 \le \log(L_\mathrm{IR}/\lsun) < 12$. The SDSS-\iras\ galaxies present a strong correlation between \sfr\ and \mstar\ (the Spearman test indicates a 18-$\sigma$ significance level for this correlation). Previous studies of local and intermediate-redshift galaxies have found that star-forming galaxies exhibit a strong correlation between stellar mass and star formation rate, forming a tight `main sequence' in which the range of possible star formation rates of a galaxy is defined by its stellar mass
\citep{Brinchmann2004,Noeske2007a,Elbaz2007}. A similar correlation has also been found for galaxies at $z\simeq1$ \citep{Elbaz2007} and at $z\simeq2$ \citep{Daddi2007}, with the overall normalization of the correlation increasing with redshift, i.e. typical galaxies present higher star formation rates for a given stellar mass at high redshifts. \cite{Noeske2007b} explain this correlation in terms of a mass-dependent gradual decline of the star formation in disk galaxies as a result of gas exhaustion as they evolve. It is interesting to analyze the distribution of local ULIRGs in the star formation vs. stellar mass plot.
The ULIRGs are clearly above the relation defined by the SDSS-\iras\ galaxies, i.e. at fixed stellar mass, ULIRGs have higher star formation rates (by about one order of magnitude). Additionally, our ULIRGs do not show any statistically significant correlation between \sfr\ and \mstar.

In Fig.~\ref{fig:mdust}(b), we plot the median-likelihood estimates of star formation rate against dust mass for the two samples. \cite{daCunha2009} find a very tight correlation between $\psi$ and \mdust\ for the SDSS-\iras\ galaxies. Interestingly, our ULIRGs do not follow this relation. For a given dust mass, the ULIRGs have higher star formation rates than the SDSS galaxies -- they are forming stars about 10 times more efficiently for a given \mdust. If we assume a constant dust-to-gas ratio and rely on the Schmidt-Kennicutt law \citep{Schmidt1959,Kennicutt1998}, this implies denser star-forming gas in ULIRGs (as observed by e.g.~\citealt{Gao2004,Bryant1999,Iono2009,Juneau2009}).
These results are consistent with the scenario in which gas and dust in local ULIRGs are concentrated in a compact central region (e.g.~\citealt{Soifer2001}; D\'iaz-Santos et al., 2010) following a merger of two gas-rich galaxies, and creating a strong starburst that powers the huge infrared luminosities of these systems (as found by simulations of major mergers of spirals; e.g.~\citealt{Mihos1994,Mihos1999,Hopkins2006}).

While these findings are not new, we show here how our SED modelling can be useful to derive relevant physical parameters which provide clues to the star formation mode of galaxies by using simple photometric observations. Additionally to a higher star formation efficiency, we also show dust in ULIRGs is typically warmer, as shown in the histograms of Fig.~\ref{fig:hist}. This is again consistent with the idea that the star formation in these galaxies is more spatially concentrated, and dust grains remain closer to the young, newly-formed stars, reaching higher equilibrium temperatures (see also, e.g.~\citealt{Yang2007,Groves2008}).

\section{Summary and Conclusions}\label{conclusion}

In this paper, we have studied the physical properties of a sample of 16 local star-forming ULIRGs observed in the mid-infrared with {\it Spitzer}/IRS for which we have compiled additional photometry from the ultraviolet to the infrared.

We have interpreted the observed ultraviolet-to-infrared spectral energy distributions of these galaxies using the simple and physically-motivated model of \cite{daCunha2008}.
Using this prescription, we describe our ULIRGs using two main components: (i) an `optically-thick starburst' with a lifetime of 100~Myr, analogous to the `birth cloud' component in the original model of \cite{daCunha2008}, where dust is heated by young stars; (ii) a diffuse ISM component, with smaller dust effective optical depths, where dust is heated by stars older than 100~Myr. The model accounts for the emission of stellar populations from the ultraviolet to the near-infrared and for the attenuation by dust in the optically-thick starburst component and in the diffuse ISM. The infrared emission is simply computed by assuming that all the energy absorbed by dust in these components is re-radiated at mid- and far-infrared wavelengths, using four main dust components.
We have combined this model with a Bayesian technique to derive statistical constraints on parameters such as the star formation rate, stellar mass, dust mass, dust temperature, and contribution of each dust component to the total infrared luminosity from their observed multi-wavelength spectral energy distributions.

We show that the model of \cite{daCunha2008} is versatile enough to successfully account for the observed spectral energy distributions of star-forming ULIRGs. Although the ultraviolet to near-infrared emission represents only a small fraction of the total power radiated by ULIRGs, observations in this wavelength range are important to constrain the dust attenuation and the fraction of infrared luminosity contributed by the diffuse ISM, the total stellar mass, and the specific star formation rate.
Additionally, our method can also provide clues to the star formation mode of our galaxies. The best fit models to the spectral energy distributions of our ULIRGs indicate that the observations can only be optimally matched if the star formation history includes a strong burst of star formation at the present age. While this is not surprising, given the known starburst nature of these galaxies, it demonstrates the consistency of our method.

To put the physical properties of local star-forming ULIRGs in context, we have compared them with a sample of SDSS galaxies of lower infrared luminosities ($10 \le \log(L_\mathrm{IR}/\lsun) < 12$) detected in the infrared with \iras\ \citep{daCunha2009}. Our results are consistent with a well-known fundamental difference in the star formation mode of local ULIRGs and other local galaxies of lower infrared luminosities. ULIRGs have star formation rates and dust temperatures significantly higher than the other galaxies, consistent with a merger-induced starburst, in which gas is driven to the central region, forming stars at very high rates and thus powering the huge infrared luminosities of these galaxies.

The method used in this paper provides a framework to better understand local star-forming galaxies spanning a wide range of infrared luminosities in a statistical manner. This has the additional advantage that it can also be used to preform quantitative comparisons to high-redshift galaxies, for which often much more sparsely sampled SEDs are available.

In future work, we plan to include the contribution by AGN in our modelling of the spectral energy distributions of galaxies, in order to account for all types of LIRGs and ULIRGs in local samples \citep{Armus2009,Desai2007}. 

\begin{acknowledgements}
We thank the anonymous referee for comments and suggestions that significantly helped improve this paper.
EdC, VC and TDS acknowledge partial support from the EU ToK grant 39965 and FP7-REGPOT 206469. We thank David Sanders and Vivian U for providing us the optical photometry for some of our galaxies. We are grateful to Bernhard Brandl and Emanuele Daddi for useful discussions, and to Brent Groves, Nick Kylafis, Vivienne Wild and Stephanie Juneau for comments on the manuscript. This research has made use of the NASA/IPAC Extragalactic Database (NED) which is operated by the Jet Propulsion Laboratory, California Institute of Technology, under contract with the National Aeronautics and Space Administration.

\end{acknowledgements}


\def\aj{AJ}
\def\araa{ARA\&A}
\def\apj{ApJ}
\def\apjl{ApJ}
\def\apjs{ApJS}
\def\apss{Ap\&SS}
\def\aap{A\&A}
\def\aapr{A\&A~Rev.}
\def\aaps{A\&AS}
\def\mnras{MNRAS}
\def\pasp{PASP}
\def\pasj{PASJ}
\def\qjras{QJRAS}
\def\nat{Nature}
\def\aplett{Astrophys.~Lett.}
\def\aas{AAS}
\let\astap=\aap
\let\apjlett=\apjl
\let\apjsupp=\apjs
\let\applopt=\ao

\bibliographystyle{aa}
\bibliography{bib_dacunha}

\begin{thebibliography}{82}
\expandafter\ifx\csname natexlab\endcsname\relax\def\natexlab#1{#1}\fi

\bibitem[{{Armus} {et~al.}(2007){Armus}, {Charmandaris}, {Bernard-Salas},
  {Spoon}, {Marshall}, {Higdon}, {Desai}, {Teplitz}, {Hao}, {Devost}, {Brandl},
  {Wu}, {Sloan}, {Soifer}, {Houck}, \& {Herter}}]{Armus2007}
{Armus}, L., {Charmandaris}, V., {Bernard-Salas}, J., {et~al.} 2007, \apj, 656,
  148

\bibitem[{{Armus} {et~al.}(2009){Armus}, {Mazzarella}, {Evans}, {Surace},
  {Sanders}, {Iwasawa}, {Frayer}, {Howell}, {Chan}, {Petric}, {Vavilkin},
  {Kim}, {Haan}, {Inami}, {Murphy}, {Appleton}, {Barnes}, {Bothun}, {Bridge},
  {Charmandaris}, {Jensen}, {Kewley}, {Lord}, {Madore}, {Marshall},
  {Melbourne}, {Rich}, {Satyapal}, {Schulz}, {Spoon}, {Sturm}, {U}, {Veilleux},
  \& {Xu}}]{Armus2009}
{Armus}, L., {Mazzarella}, J.~M., {Evans}, A.~S., {et~al.} 2009, \pasp, 121,
  559

\bibitem[{{Blain} {et~al.}(2002){Blain}, {Smail}, {Ivison}, {Kneib}, \&
  {Frayer}}]{Blain2002}
{Blain}, A.~W., {Smail}, I., {Ivison}, R.~J., {Kneib}, J.-P., \& {Frayer},
  D.~T. 2002, Physics Reports, 369, 111

\bibitem[{{Brandl} {et~al.}(2006){Brandl}, {Bernard-Salas}, {Spoon}, {Devost},
  {Sloan}, {Guilles}, {Wu}, {Houck}, {Weedman}, {Armus}, {Appleton}, {Soifer},
  {Charmandaris}, {Hao}, {Higdon}, {Marshall}, \& {Herter}}]{Brandl2006}
{Brandl}, B.~R., {Bernard-Salas}, J., {Spoon}, H.~W.~W., {et~al.} 2006, \apj,
  653, 1129

\bibitem[{{Brinchmann} {et~al.}(2004){Brinchmann}, {Charlot}, {White},
  {Tremonti}, {Kauffmann}, {Heckman}, \& {Brinkmann}}]{Brinchmann2004}
{Brinchmann}, J., {Charlot}, S., {White}, S.~D.~M., {et~al.} 2004, \mnras, 351,
  1151

\bibitem[{{Bruzual}(2007)}]{Bruzual2007}
{Bruzual}, G. 2007, (astro-ph/0703052)

\bibitem[{{Bruzual} \& {Charlot}(2003)}]{Bruzual2003}
{Bruzual}, G. \& {Charlot}, S. 2003, \mnras, 344, 1000

\bibitem[{{Bryant} \& {Scoville}(1996)}]{Bryant1996}
{Bryant}, P.~M. \& {Scoville}, N.~Z. 1996, \apj, 457, 678

\bibitem[{{Bryant} \& {Scoville}(1999)}]{Bryant1999}
{Bryant}, P.~M. \& {Scoville}, N.~Z. 1999, \aj, 117, 2632

\bibitem[{{Chabrier}(2003)}]{Chabrier2003}
{Chabrier}, G. 2003, \pasp, 115, 763

\bibitem[{{Charlot} \& {Fall}(2000)}]{Charlot2000}
{Charlot}, S. \& {Fall}, S.~M. 2000, \apj, 539, 718

\bibitem[{{Charmandaris}(2008)}]{Charmandaris2008}
{Charmandaris}, V. 2008, in Astronomical Society of the Pacific Conference
  Series, Vol. 381, Infrared Diagnostics of Galaxy Evolution, ed. {R.-R.~Chary,
  H.~I.~Teplitz, \& K.~Sheth}, 3--+

\bibitem[{{Chary} \& {Elbaz}(2001)}]{Chary2001}
{Chary}, R. \& {Elbaz}, D. 2001, \apj, 556, 562

\bibitem[{{Clements} {et~al.}(2010){Clements}, {Dunne}, \&
  {Eales}}]{Clements2010}
{Clements}, D.~L., {Dunne}, L., \& {Eales}, S. 2010, \mnras, 403, 274

\bibitem[{{Clements} {et~al.}(1996){Clements}, {Sutherland}, {McMahon}, \&
  {Saunders}}]{Clements1996}
{Clements}, D.~L., {Sutherland}, W.~J., {McMahon}, R.~G., \& {Saunders}, W.
  1996, \mnras, 279, 477

\bibitem[{{da Cunha} {et~al.}(2008){da Cunha}, {Charlot}, \&
  {Elbaz}}]{daCunha2008}
{da Cunha}, E., {Charlot}, S., \& {Elbaz}, D. 2008, \mnras, 388, 1595

\bibitem[{{da Cunha} {et~al.}(2010){da Cunha}, {Eminian}, {Charlot}, \&
  {Blaizot}}]{daCunha2009}
{da Cunha}, E., {Eminian}, C., {Charlot}, S., \& {Blaizot}, J. 2010, \mnras,
  273

\bibitem[{{Daddi} {et~al.}(2007){Daddi}, {Dickinson}, {Morrison}, {Chary},
  {Cimatti}, {Elbaz}, {Frayer}, {Renzini}, {Pope}, {Alexander}, {Bauer},
  {Giavalisco}, {Huynh}, {Kurk}, \& {Mignoli}}]{Daddi2007}
{Daddi}, E., {Dickinson}, M., {Morrison}, G., {et~al.} 2007, \apj, 670, 156

\bibitem[{{Desai} {et~al.}(2007){Desai}, {Armus}, {Spoon}, {Charmandaris},
  {Bernard-Salas}, {Brandl}, {Farrah}, {Soifer}, {Teplitz}, {Ogle}, {Devost},
  {Higdon}, {Marshall}, \& {Houck}}]{Desai2007}
{Desai}, V., {Armus}, L., {Spoon}, H.~W.~W., {et~al.} 2007, \apj, 669, 810

\bibitem[{{Dey} {et~al.}(2008){Dey}, {Soifer}, {Desai}, {Brand}, {Le Floc'h},
  {Brown}, {Jannuzi}, {Armus}, {Bussmann}, {Brodwin}, {Bian}, {Eisenhardt},
  {Higdon}, {Weedman}, \& {Willner}}]{Dey2008}
{Dey}, A., {Soifer}, B.~T., {Desai}, V., {et~al.} 2008, \apj, 677, 943

\bibitem[{{Downes} \& {Solomon}(1998)}]{Downes1998}
{Downes}, D. \& {Solomon}, P.~M. 1998, \apj, 507, 615

\bibitem[{{Duc} {et~al.}(1997){Duc}, {Mirabel}, \& {Maza}}]{Duc1997}
{Duc}, P., {Mirabel}, I.~F., \& {Maza}, J. 1997, \aaps, 124, 533

\bibitem[{{Dunne} {et~al.}(2000){Dunne}, {Eales}, {Edmunds}, {Ivison},
  {Alexander}, \& {Clements}}]{Dunne2000}
{Dunne}, L., {Eales}, S., {Edmunds}, M., {et~al.} 2000, \mnras, 315, 115

\bibitem[{{Elbaz} {et~al.}(2007){Elbaz}, {Daddi}, {Le Borgne}, {Dickinson},
  {Alexander}, {Chary}, {Starck}, {Brandt}, {Kitzbichler}, {MacDonald},
  {Nonino}, {Popesso}, {Stern}, \& {Vanzella}}]{Elbaz2007}
{Elbaz}, D., {Daddi}, E., {Le Borgne}, D., {et~al.} 2007, \aap, 468, 33

\bibitem[{{Farrah} {et~al.}(2003){Farrah}, {Afonso}, {Efstathiou},
  {Rowan-Robinson}, {Fox}, \& {Clements}}]{Farrah2003}
{Farrah}, D., {Afonso}, J., {Efstathiou}, A., {et~al.} 2003, \mnras, 343, 585

\bibitem[{{Farrah} {et~al.}(2007){Farrah}, {Bernard-Salas}, {Spoon}, {Soifer},
  {Armus}, {Brandl}, {Charmandaris}, {Desai}, {Higdon}, {Devost}, \&
  {Houck}}]{Farrah2007}
{Farrah}, D., {Bernard-Salas}, J., {Spoon}, H.~W.~W., {et~al.} 2007, \apj, 667,
  149

\bibitem[{{Gao} \& {Solomon}(2004)}]{Gao2004}
{Gao}, Y. \& {Solomon}, P.~M. 2004, \apj, 606, 271

\bibitem[{{Garc{\'{\i}}a-Mar{\'{\i}}n}
  {et~al.}(2009){Garc{\'{\i}}a-Mar{\'{\i}}n}, {Colina}, \&
  {Arribas}}]{Garcia2009}
{Garc{\'{\i}}a-Mar{\'{\i}}n}, M., {Colina}, L., \& {Arribas}, S. 2009, \aap,
  505, 1017

\bibitem[{{Genzel} {et~al.}(1998){Genzel}, {Lutz}, {Sturm}, {Egami}, {Kunze},
  {Moorwood}, {Rigopoulou}, {Spoon}, {Sternberg}, {Tacconi-Garman}, {Tacconi},
  \& {Thatte}}]{Genzel1998}
{Genzel}, R., {Lutz}, D., {Sturm}, E., {et~al.} 1998, \apj, 498, 579

\bibitem[{{Genzel} {et~al.}(2001){Genzel}, {Tacconi}, {Rigopoulou}, {Lutz}, \&
  {Tecza}}]{Genzel2001}
{Genzel}, R., {Tacconi}, L.~J., {Rigopoulou}, D., {Lutz}, D., \& {Tecza}, M.
  2001, \apj, 563, 527

\bibitem[{{Groves} {et~al.}(2008){Groves}, {Dopita}, {Sutherland}, {Kewley},
  {Fischera}, {Leitherer}, {Brandl}, \& {van Breugel}}]{Groves2008}
{Groves}, B., {Dopita}, M.~A., {Sutherland}, R.~S., {et~al.} 2008, \apjs, 176,
  438

\bibitem[{{Hao} {et~al.}(2007){Hao}, {Weedman}, {Spoon}, {Marshall},
  {Levenson}, {Elitzur}, \& {Houck}}]{Hao2007}
{Hao}, L., {Weedman}, D.~W., {Spoon}, H.~W.~W., {et~al.} 2007, \apjl, 655, L77

\bibitem[{{Hopkins} {et~al.}(2006){Hopkins}, {Hernquist}, {Cox}, {Di Matteo},
  {Robertson}, \& {Springel}}]{Hopkins2006}
{Hopkins}, P.~F., {Hernquist}, L., {Cox}, T.~J., {et~al.} 2006, \apjs, 163, 1

\bibitem[{{Imanishi} {et~al.}(2006){Imanishi}, {Dudley}, \&
  {Maloney}}]{Imanishi2006}
{Imanishi}, M., {Dudley}, C.~C., \& {Maloney}, P.~R. 2006, \apj, 637, 114

\bibitem[{{Iono} {et~al.}(2009){Iono}, {Wilson}, {Yun}, {Baker}, {Petitpas},
  {Peck}, {Krips}, {Cox}, {Matsushita}, {Mihos}, \& {Pihlstrom}}]{Iono2009}
{Iono}, D., {Wilson}, C.~D., {Yun}, M.~S., {et~al.} 2009, \apj, 695, 1537

\bibitem[{{Juneau} {et~al.}(2009){Juneau}, {Narayanan}, {Moustakas}, {Shirley},
  {Bussmann}, {Kennicutt}, \& {Vanden Bout}}]{Juneau2009}
{Juneau}, S., {Narayanan}, D.~T., {Moustakas}, J., {et~al.} 2009, \apj, 707,
  1217

\bibitem[{{Kauffmann et al.}(2003)}]{Kauffmann2003b}
{Kauffmann et al.} 2003, \mnras, 341, 54

\bibitem[{{Kennicutt}(1998)}]{Kennicutt1998}
{Kennicutt}, Jr., R.~C. 1998, \apj, 498, 541

\bibitem[{{Kennicutt} {et~al.}(2003){Kennicutt}, {Armus}, {Bendo}, {Calzetti},
  {Dale}, {Draine}, {Engelbracht}, {Gordon}, \& {et al.}}]{Kennicutt2003}
{Kennicutt}, Jr., R.~C., {Armus}, L., {Bendo}, G., {et~al.} 2003, \pasp, 115,
  928

\bibitem[{{Kim} \& {Sanders}(1998)}]{Kim1998}
{Kim}, D. \& {Sanders}, D.~B. 1998, \apjs, 119, 41

\bibitem[{{Kim} {et~al.}(1998){Kim}, {Veilleux}, \& {Sanders}}]{Kimetal1998}
{Kim}, D., {Veilleux}, S., \& {Sanders}, D.~B. 1998, \apj, 508, 627

\bibitem[{{Klaas} {et~al.}(2001){Klaas}, {Haas}, {M{\"u}ller}, {Chini},
  {Schulz}, {Coulson}, {Hippelein}, {Wilke}, {Albrecht}, \&
  {Lemke}}]{Klaas2001}
{Klaas}, U., {Haas}, M., {M{\"u}ller}, S.~A.~H., {et~al.} 2001, \aap, 379, 823

\bibitem[{{Lahuis} {et~al.}(2007){Lahuis}, {Spoon}, {Tielens}, {Doty}, {Armus},
  {Charmandaris}, {Houck}, {St{\"a}uber}, \& {van Dishoeck}}]{Lahuis2007}
{Lahuis}, F., {Spoon}, H.~W.~W., {Tielens}, A.~G.~G.~M., {et~al.} 2007, \apj,
  659, 296

\bibitem[{{Laurent} {et~al.}(2000){Laurent}, {Mirabel}, {Charmandaris},
  {Gallais}, {Madden}, {Sauvage}, {Vigroux}, \& {Cesarsky}}]{Laurent2000}
{Laurent}, O., {Mirabel}, I.~F., {Charmandaris}, V., {et~al.} 2000, \aap, 359,
  887

\bibitem[{{Le Floc'h} {et~al.}(2005){Le Floc'h}, {Papovich}, {Dole}, \& {et
  al.}}]{LeFloch2005}
{Le Floc'h}, E., {Papovich}, C., {Dole}, H., \& {et al.} 2005, \apj, 632, 169

\bibitem[{{Lutz} {et~al.}(1998){Lutz}, {Spoon}, {Rigopoulou}, {Moorwood}, \&
  {Genzel}}]{Lutz1998}
{Lutz}, D., {Spoon}, H.~W.~W., {Rigopoulou}, D., {Moorwood}, A.~F.~M., \&
  {Genzel}, R. 1998, \apjl, 505, L103

\bibitem[{{Lutz} {et~al.}(1999){Lutz}, {Veilleux}, \& {Genzel}}]{Lutz1999}
{Lutz}, D., {Veilleux}, S., \& {Genzel}, R. 1999, \apjl, 517, L13

\bibitem[{{Maddox} {et~al.}(1990){Maddox}, {Efstathiou}, {Sutherland}, \&
  {Loveday}}]{Maddox1990}
{Maddox}, S.~J., {Efstathiou}, G., {Sutherland}, W.~J., \& {Loveday}, J. 1990,
  \mnras, 243, 692

\bibitem[{{Marshall} {et~al.}(2007){Marshall}, {Herter}, {Armus},
  {Charmandaris}, {Spoon}, {Bernard-Salas}, \& {Houck}}]{Marshall2007}
{Marshall}, J.~A., {Herter}, T.~L., {Armus}, L., {et~al.} 2007, \apj, 670, 129

\bibitem[{{Mathis} {et~al.}(1977){Mathis}, {Rumpl}, \&
  {Nordsieck}}]{Mathis1977}
{Mathis}, J.~S., {Rumpl}, W., \& {Nordsieck}, K.~H. 1977, \apj, 217, 425

\bibitem[{{Mihos}(1999)}]{Mihos1999}
{Mihos}, J.~C. 1999, in IAU Symposium, Vol. 186, Galaxy Interactions at Low and
  High Redshift, ed. {J.~E.~Barnes \& D.~B.~Sanders}, 205--+

\bibitem[{{Mihos} \& {Hernquist}(1994)}]{Mihos1994}
{Mihos}, J.~C. \& {Hernquist}, L. 1994, \apjl, 431, L9

\bibitem[{{Moshir} \& {et al.}(1990)}]{Moshir1990}
{Moshir}, M. \& {et al.} 1990, in IRAS Faint Source Catalogue, version 2.0
  (1990), 0--+

\bibitem[{{Murphy} {et~al.}(2001){Murphy}, {Soifer}, {Matthews}, \&
  {Armus}}]{Murphy2001}
{Murphy}, Jr., T.~W., {Soifer}, B.~T., {Matthews}, K., \& {Armus}, L. 2001,
  \apj, 559, 201

\bibitem[{{Noeske} {et~al.}(2007{\natexlab{a}}){Noeske}, {Faber}, {Weiner},
  {Koo}, {Primack}, {Dekel}, {Papovich}, {Conselice}, {Le Floc'h}, {Rieke},
  {Coil}, {Lotz}, {Somerville}, \& {Bundy}}]{Noeske2007b}
{Noeske}, K.~G., {Faber}, S.~M., {Weiner}, B.~J., {et~al.} 2007{\natexlab{a}},
  \apjl, 660, L47

\bibitem[{{Noeske} {et~al.}(2007{\natexlab{b}}){Noeske}, {Weiner}, {Faber},
  {Papovich}, {Koo}, {Somerville}, {Bundy}, {Conselice}, {Newman},
  {Schiminovich}, {Le Floc'h}, {Coil}, {Rieke}, {Lotz}, {Primack}, {Barmby},
  {Cooper}, {Davis}, {Ellis}, {Fazio}, {Guhathakurta}, {Huang}, {Kassin},
  {Martin}, {Phillips}, {Rich}, {Small}, {Willmer}, \& {Wilson}}]{Noeske2007a}
{Noeske}, K.~G., {Weiner}, B.~J., {Faber}, S.~M., {et~al.} 2007{\natexlab{b}},
  \apjl, 660, L43

\bibitem[{{Ossenkopf} {et~al.}(1992){Ossenkopf}, {Henning}, \&
  {Mathis}}]{Ossenkopf1992}
{Ossenkopf}, V., {Henning}, T., \& {Mathis}, J.~S. 1992, \aap, 261, 567

\bibitem[{{Rigopoulou} {et~al.}(1996){Rigopoulou}, {Lawrence}, \&
  {Rowan-Robinson}}]{Rigopoulou1996}
{Rigopoulou}, D., {Lawrence}, A., \& {Rowan-Robinson}, M. 1996, \mnras, 278,
  1049

\bibitem[{{Rigopoulou et al.}(1999)}]{Rigopoulou1999}
{Rigopoulou et al.} 1999, in ESA Special Publication, Vol. 427, The Universe as
  Seen by ISO, ed. P.~{Cox} \& M.~{Kessler}, 833--+

\bibitem[{{Rodr{\'{\i}}guez Zaur{\'{\i}}n} {et~al.}(2008){Rodr{\'{\i}}guez
  Zaur{\'{\i}}n}, {Tadhunter}, \& {Gonz{\'a}lez Delgado}}]{Rodriguez2008}
{Rodr{\'{\i}}guez Zaur{\'{\i}}n}, J., {Tadhunter}, C.~N., \& {Gonz{\'a}lez
  Delgado}, R.~M. 2008, \mnras, 384, 875

\bibitem[{{Rodr{\'{\i}}guez Zaur{\'{\i}}n} {et~al.}(2010){Rodr{\'{\i}}guez
  Zaur{\'{\i}}n}, {Tadhunter}, \& {Gonz{\'a}lez Delgado}}]{Rodriguez2010}
{Rodr{\'{\i}}guez Zaur{\'{\i}}n}, J., {Tadhunter}, C.~N., \& {Gonz{\'a}lez
  Delgado}, R.~M. 2010, \mnras, 403, 1317

\bibitem[{{Sakamoto} {et~al.}(2008){Sakamoto}, {Wang}, {Wiedner}, {Wang},
  {Peck}, {Zhang}, {Petitpas}, {Ho}, \& {Wilner}}]{Sakamoto2008}
{Sakamoto}, K., {Wang}, J., {Wiedner}, M.~C., {et~al.} 2008, \apj, 684, 957

\bibitem[{{Salpeter}(1955)}]{Salpeter1955}
{Salpeter}, E.~E. 1955, \apj, 121, 161

\bibitem[{{Sanders} {et~al.}(2003){Sanders}, {Mazzarella}, {Kim}, {Surace}, \&
  {Soifer}}]{Sanders2003}
{Sanders}, D.~B., {Mazzarella}, J.~M., {Kim}, D., {Surace}, J.~A., \& {Soifer},
  B.~T. 2003, \aj, 126, 1607

\bibitem[{{Sanders} \& {Mirabel}(1996)}]{Sanders1996}
{Sanders}, D.~B. \& {Mirabel}, I.~F. 1996, \araa, 34, 749

\bibitem[{{Sanders} {et~al.}(1988){Sanders}, {Soifer}, {Elias}, {Neugebauer},
  \& {Matthews}}]{Sanders1988}
{Sanders}, D.~B., {Soifer}, B.~T., {Elias}, J.~H., {Neugebauer}, G., \&
  {Matthews}, K. 1988, \apjl, 328, L35

\bibitem[{{Schmidt}(1959)}]{Schmidt1959}
{Schmidt}, M. 1959, \apj, 129, 243

\bibitem[{{Scoville} {et~al.}(2000){Scoville}, {Evans}, {Thompson}, {Rieke},
  {Hines}, {Low}, {Dinshaw}, {Surace}, \& {Armus}}]{Scoville2000}
{Scoville}, N.~Z., {Evans}, A.~S., {Thompson}, R., {et~al.} 2000, \aj, 119, 991

\bibitem[{{Sirocky} {et~al.}(2008){Sirocky}, {Levenson}, {Elitzur}, {Spoon}, \&
  {Armus}}]{Sirocky2008}
{Sirocky}, M.~M., {Levenson}, N.~A., {Elitzur}, M., {Spoon}, H.~W.~W., \&
  {Armus}, L. 2008, \apj, 678, 729

\bibitem[{{Soifer} {et~al.}(2001){Soifer}, {Neugebauer}, {Matthews}, {Egami},
  {Weinberger}, {Ressler}, {Scoville}, {Stolovy}, {Condon}, \&
  {Becklin}}]{Soifer2001}
{Soifer}, B.~T., {Neugebauer}, G., {Matthews}, K., {et~al.} 2001, \aj, 122,
  1213

\bibitem[{{Soto} \& {Martin}(2010)}]{Soto2010}
{Soto}, K.~T. \& {Martin}, C.~L. 2010, \apj, 716, 332

\bibitem[{{Spoon} {et~al.}(2007){Spoon}, {Marshall}, {Houck}, {Elitzur}, {Hao},
  {Armus}, {Brandl}, \& {Charmandaris}}]{Spoon2007}
{Spoon}, H.~W.~W., {Marshall}, J.~A., {Houck}, J.~R., {et~al.} 2007, \apjl,
  654, L49

\bibitem[{{Stanford} {et~al.}(2000){Stanford}, {Stern}, {van Breugel}, \& {De
  Breuck}}]{Stanford2000}
{Stanford}, S.~A., {Stern}, D., {van Breugel}, W., \& {De Breuck}, C. 2000,
  \apjs, 131, 185

\bibitem[{{Stickel} {et~al.}(2004){Stickel}, {Lemke}, {Klaas}, {Krause}, \&
  {Egner}}]{Stickel2004}
{Stickel}, M., {Lemke}, D., {Klaas}, U., {Krause}, O., \& {Egner}, S. 2004,
  \aap, 422, 39

\bibitem[{{Strauss} {et~al.}(1992){Strauss}, {Huchra}, {Davis}, {Yahil},
  {Fisher}, \& {Tonry}}]{Strauss1992}
{Strauss}, M.~A., {Huchra}, J.~P., {Davis}, M., {et~al.} 1992, \apjs, 83, 29

\bibitem[{{Surace} \& {Sanders}(2000)}]{Surace2000}
{Surace}, J.~A. \& {Sanders}, D.~B. 2000, \aj, 120, 604

\bibitem[{{Surace} {et~al.}(2000){Surace}, {Sanders}, \& {Evans}}]{Surace2000b}
{Surace}, J.~A., {Sanders}, D.~B., \& {Evans}, A.~S. 2000, \apj, 529, 170

\bibitem[{{Tacconi} {et~al.}(2002){Tacconi}, {Genzel}, {Lutz}, {Rigopoulou},
  {Baker}, {Iserlohe}, \& {Tecza}}]{Tacconi2002}
{Tacconi}, L.~J., {Genzel}, R., {Lutz}, D., {et~al.} 2002, \apj, 580, 73

\bibitem[{{Vega} {et~al.}(2008){Vega}, {Clemens}, {Bressan}, {Granato},
  {Silva}, \& {Panuzzo}}]{Vega2008}
{Vega}, O., {Clemens}, M.~S., {Bressan}, A., {et~al.} 2008, \aap, 484, 631

\bibitem[{{Veilleux} {et~al.}(2002){Veilleux}, {Kim}, \&
  {Sanders}}]{Veilleux2002}
{Veilleux}, S., {Kim}, D., \& {Sanders}, D.~B. 2002, \apjs, 143, 315

\bibitem[{{Veilleux} {et~al.}(2009){Veilleux}, {Rupke}, {Kim}, {Genzel},
  {Sturm}, {Lutz}, {Contursi}, {Schweitzer}, {Tacconi}, {Netzer}, {Sternberg},
  {Mihos}, {Baker}, {Mazzarella}, {Lord}, {Sanders}, {Stockton}, {Joseph}, \&
  {Barnes}}]{Veilleux2009}
{Veilleux}, S., {Rupke}, D.~S.~N., {Kim}, D., {et~al.} 2009, \apjs, 182, 628

\bibitem[{{Yang} {et~al.}(2007){Yang}, {Greve}, {Dowell}, \&
  {Borys}}]{Yang2007}
{Yang}, M., {Greve}, T.~R., {Dowell}, C.~D., \& {Borys}, C. 2007, \apj, 660,
  1198

\end{thebibliography}

\begin{appendix}
\section{Multi-wavelength photometric data}\label{data}
\begin{sidewaystable*}
\vspace{12cm}
\caption{Photometric data -- ultraviolet and optical. All fluxes are in Jy.} \label{tab:data_opt}
\begin{center}
\begin{tabular}{lccccccccccccccccccc} \hline \hline 
\\
Galaxy & $z$ & FUV & NUV & U & B & V & R & I & $u$ & $g$ & $r$ & $i$ & $z$ \\
\\
 \hline
IRAS00199-7426 & 0.0963 & - & - & - & - & - & - & - & - & - & - & - & - \\
IRAS01494-1845 & 0.1579 & - & - & - & 1.09E-4$^{(a)}$ & - & - & - & - & - & - & - & -\\
IRASZ02376-0054 & 0.4097 & - & - & - & - & - & - & - & 7.91E-7$^{(b)}$ & 5.43E-6$^{(b)}$ & 2.46E-5$^{(b)}$ & 4.57E-5$^{(b)}$ & 7.05E-5$^{(b)}$\\
IRAS04114-5117 & 0.1245 & - & - & - & - & - & - & - & - & - & - & - & -\\
IRAS06009-7716 & 0.1169 & - & - & - & - & - & - & - & - & - & - & - & -\\
IRASF08208+3211 & 0.3957 & - & - & - & - & - & - & - & 1.04E-5$^{(b)}$ & 2.33E-5$^{(b)}$ & 5.8E-5$^{(b)}$ & 9.92E-5$^{(b)}$ & 1.22E-4$^{(b)}$\\
IRASF10156+3705 & 0.4897 & - & - & - & - & - & - & - & - & - & - & - & -\\
IRAS10565+2448 & 0.0431 & - & - & - & - & - & - & - & 5.34E-4$^{(b)}$ & 2.01E-3$^{(b)}$ & 4.13E-3$^{(b)}$ & 6.57E-3$^{(b)}$ & 9.18E-3$^{(b)}$\\
IRAS12112+0305 & 0.0726 & 1.0E-4$^{(c)}$ & 2.12E-4$^{(c)}$ & 4.55E-4$^{(d)}$ & 6.75E-4$^{(d)}$ & - & - & - & - & - & - & - & -\\
IRAS13120-5453 & 0.0307 & - & - & - & - & - & - & - & - & - & - & - & -\\
IRAS16334+4630 & 0.1909 & - & - & - & - & - & - & - & - & - & - & - & -\\
IRAS17208-0014 & 0.043 & - & - & - & 5.25E-4$^{(e)}$ & - & 2.99E-3$^{(e)}$ & - & - & - & - & - & -\\
IRAS19297-0406 & 0.0857 & - & - & - & 6.67E-5$^{(e)}$ & - & 7.52E-4$^{(e)}$ & - & - & - & - & - & -\\
IRAS19458+0944 & 0.0999 & - & - & - & - & - & - & - & - & - & - & - & -\\
IRAS20414-1651 & 0.087 & - & - & - & 1.55E-4$^{(f)}$ & - & - & 5.84E-4$^{(f)}$ & - & - & - & - & -\\
IRAS22491-1808 & 0.0772 & - & - & 2.19E-4$^{(e)}$ & 1.06E-3$^{(e)}$ & - & 2.09E-3$^{(e)}$ & 2.29E-3$^{(e)}$ & - & - & - & - & -\\
\hline
\end{tabular}
\end{center}
\textsc{Notes:} $(a)$ -- SDSS Data Release 6 (model magnitudes); $(b)$ -- \cite{Maddox1990}; $(c)$ -- Howell et al. (2010); $(d)$ -- \cite{Surace2000};  $(e)$ -- D. Sanders and V. U, private communication; $(f)$ -- \cite{Surace2000b}. \\
\end{sidewaystable*}

\begin{sidewaystable*}
\vspace{12cm}
\caption{Photometric data -- near-infrared to mid-infrared. All fluxes are in Jy.} \label{tab:data_nir}
\begin{center}
\begin{tabular}{lccccccccccc} \hline \hline
\\
Galaxy & $z$  & J & H & K$_s$ & F110 & F160 & F222 & IRAC3.6\mic\ & IRAC4.5\mic\ & IRAC5.8\mic$^a$ & IRAC8.0\mic$^a$  \\
\\
 \hline
IRAS00199-7426 & 0.0963 & 2.84E-3$^{(a)}$ & 4.09E-3$^{(a)}$ & 6.05E-3$^{(a)}$ & - & - & - & - & - & 1.09E-2$^{(b)}$ & 6.18E-2$^{(b)}$\\
IRAS01494-1845 & 0.1579 & 6.95E-4$^{(a)}$ & 9.08E-4$^{(a)}$ & 1.75E-3$^{(a)}$ & - & - & - & - & - & 2.68E-3$^{(b)}$ & 1.51E-2$^{(b)}$ \\
IRASZ02376-0054 & 0.4097 & - & - & - & - & - & - & 2.66E-4$^{(c)}$ & - & - & - \\
IRAS04114-5117 & 0.1245 & 3.47E-4$^{(a)}$ & 1.30E-3$^{(a)}$ & 1.49E-3$^{(a)}$ & - & - & - & 8.17E-4$^{(c)}$ & - & 2.61E-3$^{(b)}$ & 1.31E-2$^{(b)}$ \\
IRAS06009-7716 & 0.1169 & 1.17E-3$^{(a)}$ & 7.99E-4$^{(a)}$ & 2.16E-3$^{(a)}$ & - & - & - & - & - & 4.15E-3$^{(b)}$ & 2.68E-2$^{(b)}$ \\
IRASF08208+3211 & 0.3957 & - & - & - & - & - & - & 1.67E-4$^{(c)}$ & - & - & 9.57E-4$^{(b)}$ \\
IRASF10156+3705 & 0.4897 & - & - & - & - & - & - & 2.10E-4$^{(c)}$ & - & - & 5.30E-4$^{(b)}$ \\
IRAS10565+2448 & 0.0431 & 7.69E-3$^{(a)}$ & 1.16E-2$^{(a)}$ & 1.41E-2$^{(a)}$ & 6.21E-3$^{(d)}$ & 1.21E-2$^{(d)}$ & 1.53E-2$^{(d)}$ & 1.21E-2$^{(c)}$ & 1.26E-2$^{(c)}$ & 3.44E-2$^{(b)}$ & 1.77E-1$^{(b)}$\\
IRAS12112+0305 & 0.0726 & - & - & - & 8.98E-4$^{(d)}$ & 1.85E-3$^{(d)}$ & 2.17E-3$^{(d)}$ & 2.33E-3$^{(c)}$ & 2.70E-3$^{(c)}$ & 9.64E-3$^{(b)}$ & 5.28E-2$^{(b)}$ \\
IRAS13120-5453 & 0.0307 & 1.42E-2$^{(a)}$ & 2.51E-2$^{(a)}$ & 3.84E-2$^{(a)}$ & - & - & - & 2.95E-2$^{(c)}$ & 3.21E-2$^{(c)}$ & 1.01E-1$^{(b)}$ & 4.03E-1$^{(b)}$ \\
IRAS16334+4630 & 0.1909 & 6.19E-4$^{(a)}$ & 8.44E-4$^{(a)}$ & 1.43E-3$^{(a)}$ & - & - & - & 1.21E-3$^{(c)}$ & - & 2.76E-3$^{(b)}$ & 1.21E-2$^{(b)}$ \\
IRAS17208-0014 & 0.043 & 1.28E-2$^{(e)}$ & 1.85E-2$^{(e)}$ & 2.08E-2$^{(e)}$ & - & - & - & 1.07E-2$^{(c)}$ & 1.12E-2$^{(c)}$ & 4.60E-2$^{(b)}$ & 1.92E-1$^{(b)}$\\
IRAS19297-0406 & 0.0857 & 5.65E-3$^{(a)}$ & 5.70E-3$^{(a)}$ & 4.16E-3$^{(a)}$ & - & - & - & 3.67E-3$^{(c)}$ & 4.29E-3$^{(c)}$ & 1.02E-2$^{(b)}$ & 5.22E-2$^{(b)}$\\
IRAS19458+0944 & 0.0999 & 1.51E-3$^{(a)}$ & 2.22E-3$^{(a)}$ & 3.13E-3$^{(a)}$ & - & - & - & 2.59E-3$^{(c)}$ & - & 5.18E-3$^{(b)}$ & 3.56E-2$^{(b)}$\\
IRAS20414-1651 & 0.087 & 1.04E-3$^{(a)}$ & 1.53E-3$^{(a)}$ & 2.34E-3$^{(a)}$ & - & - & - & 1.72E-3$^{(c)}$ & - & 4.55E-3$^{(b)}$ & 2.40E-2$^{(b)}$ \\
IRAS22491-1808 & 0.0772 & - & - & - & 1.63E-3$^{(d)}$ & 2.57E-3$^{(d)}$ & 2.45E-3$^{(d)}$ & 1.88E-3$^{(c)}$ & 1.59E-3$^{(c)}$ & 4.57E-3$^{(b)}$ & 3.22E-2$^{(b)}$ \\
\hline
\end{tabular}
\end{center}
\textsc{Notes:} $(a)$ -- 2MASS Point Source Catalog; $(b)$ -- computed from convolving the observed IRS spectra with the corresponding IRAC bandpass; $(c)$ -- ; $(d)$ -- \cite{Scoville2000}; $(e)$ -- 2MASS XSC. \\
\end{sidewaystable*}

\begin{sidewaystable*}
\vspace{12cm}
\caption{Photometric data -- mid-infrared to far-infrared. All fluxes are in Jy.} \label{tab:data_fir}
\begin{center}
\begin{tabular}{lcccccccccccc} \hline \hline
\\
Galaxy & $z$  & 12\mic\ & 25\mic\ & 60\mic\ & 100\mic\ & 15\mic\ & 24\mic\ & 70\mic\ & 160\mic\ & 350\mic\ & 450\mic\ & 850\mic\  \\
\\
 \hline
IRAS00199-7426 & 0.0963 & 6.59E-2$^{(a)}$ & 3.26E-1$^{(a)}$ & 4.16E0$^{(a)}$ & 6.42E0$^{(a)}$ & 9.00E-2$^{(b)}$& - & - & - & - & - & - \\
IRAS01494-1845 & 0.1579 & 1.72E-2$^{(a)}$ & 7.58E-2$^{(a)}$ & 1.29E0$^{(a)}$ & 1.85E0$^{(a)}$ & 2.10E-2$^{(b)}$& 6.67E-2$^{(b)}$ & - & - & - & - & - \\
IRASZ02376-0054 & 0.4097 & 1.85E-3$^{(b)}$ & 4.76E-3$^{(b)}$ & 2.00E-1$^{(c)}$ & 6.90E-1$^{(c)}$ & 1.34E-2$^{(b)}$& 4.38E-3$^{(b)}$ & - & - & 6.65E-2$^{(c)}$ & - & -\\
IRAS04114-5117 & 0.1245 & 1.32E-2$^{(a)}$ & 7.40E-2$^{(a)}$ & 2.09E0$^{(a)}$ & 3.23E0$^{(a)}$ & 1.72E-2$^{(b)}$& 7.14E-2$^{(b)}$ & 2.77E0$^{(d)}$ & 1.65E0$^{(d)}$ & - & - & -\\
IRAS06009-7716 & 0.1169 & 2.75E-2$^{(a)}$ & 1.18E-1$^{(a)}$ & 1.42E0$^{(a)}$ & 2.11E0$^{(a)}$ & 3.48E-2$^{(b)}$ & 9.54E-2$^{(b)}$ & 1.63E0$^{(d)}$ & 8.46E-1$^{(d)}$ & - & - & -\\
IRASF08208+3211 & 0.3957 & 1.76E-3$^{(b)}$ & 5.50E-3$^{(b)}$ & 2.27E-1$^{(a)}$ & 5.20E-1$^{(c)}$ & 1.49E-3$^{(b)}$ & 4.98E-3$^{(b)}$ & - & 3.62E-1$^{(d)}$ & 7.60E-2$^{(c)}$ & - & -\\
IRASF10156+3705 & 0.4897 & 1.62E-3$^{(b)}$ & 3.34E-3$^{(b)}$ & 2.06E-1$^{(a)}$ & 4.71E-1$^{(a)}$ & 1.49E-3$^{(b)}$& 2.96E-3$^{(b)}$ & - & 3.66E-1$^{(d)}$ & 7.86E-2$^{(c)}$ & - & -\\
IRAS10565+2448 & 0.0431 & 2.00E-1$^{(e)}$ & 1.27E0$^{(e)}$ & 1.21E1$^{(e)}$ & 1.50E1$^{(e)}$ & 2.99E-1$^{(b)}$& 9.86E-1$^{(d)}$ & 9.59E0$^{(d)}$ & 6.66E0$^{(d)}$ & 1.24E0$^{(c)}$ & - & 6.10E-2$^{(f)}$\\
IRAS12112+0305 & 0.0726 & 5.93E-2$^{(b)}$ & 4.68E-1$^{(b)}$ & 8.18E0$^{(e)}$ & 9.46E0$^{(e)}$ & 9.30E-2$^{(b)}$ & 3.50E-1$^{(b)}$ & - & - & - & 2.81E-1$^{(g)}$ & 4.90E-2$^{(f)}$\\
IRAS13120-5453 & 0.0307 & 4.40E-1$^{(e)}$ & 2.98E0$^{(e)}$ & 4.11E1$^{(e)}$ & 5.23E1$^{(e)}$ & 6.42E-1$^{(b)}$& 2.70E0$^{(b)}$ & 3.13E1$^{(d)}$ & - & - & - & -\\
IRAS16334+4630 & 0.1909 & 1.58E-2$^{(b)}$ & 7.99E-2$^{(b)}$ & 1.19E0$^{(a)}$ & 2.09E0$^{(a)}$ & 2.10E-2$^{(b)}$ & 6.90E-2$^{(b)}$ & 1.57E0$^{(d)}$ & 9.40E-1$^{(d)}$ & - & - & -\\
IRAS17208-0014 & 0.043 & 1.78E-1$^{(b)}$ & 1.61E0$^{(e)}$ & 3.21E1$^{(e)}$ & 3.61E1$^{(e)}$ & 2.55E-1$^{(b)}$ & 1.21E0$^{(d)}$ & 1.91E1$^{(d)}$ & - & - & 1.07E0$^{(h)}$ & 1.55E-1$^{(h)}$\\
IRAS19297-0406 & 0.0857 & 6.02E-2$^{(b)}$ & 5.17E-1$^{(b)}$ & 7.32E0$^{(e)}$ & 8.62E0$^{(e)}$ & 9.87E-2$^{(b)}$ & 3.93E-1$^{(d)}$ & 6.64E0$^{(d)}$ & 4.47E0$^{(d)}$ & - & - & -\\
IRAS19458+0944 & 0.0999 & 3.73E-2$^{(b)}$ & 2.53E-1$^{(b)}$ & 3.95E0$^{(i)}$ & 7.11E0$^{(i)}$ & 5.46E-2$^{(b)}$ & 1.70E-1$^{(d)}$ & 3.32E0$^{(d)}$ & - & - & - & -\\
IRAS20414-1651 & 0.087 & 2.51E-2$^{(b)}$ & 2.64E-1$^{(b)}$ & 4.36E0$^{(a)}$ & 5.25E0$^{(a)}$ & 3.89E-2$^{(b)}$ & 2.20E-1$^{(b)}$ & - & - & - & - & -\\
IRAS22491-1808 & 0.0772 & 4.37E-2$^{(b)}$ & 5.40E-1$^{(e)}$ & 5.54E0$^{(e)}$ & 4.64E0$^{(e)}$ & 7.90E-2$^{(b)}$& 4.09E-1$^{(d)}$ & 4.68E0$^{(d)}$ & 2.04E0$^{(d)}$ & - & - & 1.90E-2$^{(h)}$\\
\hline
\end{tabular}
\end{center}
\textsc{Notes:} $(a)$ -- IRAS FSC \citep{Moshir1990}; $(b)$ -- Computed from IRS spectrum; $(c)$ -- \cite{Yang2007}; $(d)$ -- Computed using post-BCD products provided by the {\it Spitzer} Science Center, as described in \cite{Marshall2007}; $(e)$ -- \cite{Sanders2003}; $(f)$ -- \cite{Dunne2000}; $(g)$ -- \cite{Rigopoulou1996}; $(h)$ -- \cite{Klaas2001}; $(i)$ -- IRAS PSC.\\
\textsc{Extra ISO photometry:} IRAS00199-7426 -- $F_\nu(10\mic)=5.70E-1$, $F_\nu(120\mic)=7.50E0$, $F_\nu(150\mic)=5.34E0$, $F_\nu(180\mic)=4.21E0$, $F_\nu(200\mic)=3.00E0$ \citep{Klaas2001}; IRAS01494-1845 -- $F_\nu(170\mic)=1.49E0$ \citep{Stickel2004}; IRAS13120-5453 -- $F_\nu(170\mic)=3.05E1$ \citep{Stickel2004}.\\
\end{sidewaystable*}

\end{appendix}

\end{document}